\documentclass[aps,prb,10pt,superscriptaddress,twocolumn]{revtex4-2}

\usepackage{graphicx}
\usepackage{dcolumn}
\usepackage{color}
\usepackage{amsmath}
\usepackage{nicefrac}
\usepackage{amssymb}
\usepackage{bm,dsfont}
\usepackage{multirow}
\usepackage{xcolor}
\usepackage[pdftex,colorlinks=true,linkcolor=blue,citecolor=magenta,filecolor=blue]{hyperref}
\usepackage{comment}

\newcommand{\ff}[1]{{\boldsymbol #1}}
\newcommand{\ca}[1]{{\cal #1}}
\newcommand{\bi}{\begin{itemize}}
\newcommand{\ei}{\end{itemize}}
\newcommand{\be}{\begin{equation}}
\newcommand{\ee}{\end{equation}}
\newcommand{\ba}{\begin{eqnarray}}
\newcommand{\ea}{\end{eqnarray}}

\newcommand{\refeq}[1]{Eq.\ (\ref{eq:#1})}
\newcommand{\labeq}[1]{\label{eq:#1}}

\begin{document}

\title{
Disorder-Induced Topological Phases in a Two-Dimensional Chern Insulator with Strong Magnetic Disorder
}

\author{Devesh Vaish}

\affiliation{I.\ Institute of Theoretical Physics, Department of Physics, University of Hamburg, Notkestra\ss{}e 9, 22607 Hamburg, Germany}

\author{Michael Potthoff}

\affiliation{I.\ Institute of Theoretical Physics, Department of Physics, University of Hamburg, Notkestra\ss{}e 9, 22607 Hamburg, Germany}

\affiliation{The Hamburg Centre for Ultrafast Imaging, Luruper Chaussee 149, 22761 Hamburg, Germany}

\date{\today}

\begin{abstract}
Strong directional disorder in local magnetic moments coupled to a Chern insulator gives rise to topological phases that cannot be realized in any clean system and are therefore genuinely disorder-driven. We demonstrate this in a spinful Qi-Wu-Zhang model of a two-dimensional Chern insulator coupled to disordered classical spins of unit length. The topological phase diagram is computed numerically using two complementary approaches: twisted boundary conditions and the topological Hamiltonian technique. Our results show that strong disorder can act as a fundamental topological mechanism rather than merely a perturbation.
For strong exchange coupling, tuning the mass parameter reveals a transition between phases with different Chern numbers $C$. Remarkably, this transition is driven by zeros, rather than poles, of the disorder-averaged Green's function crossing the chemical potential, and has no analogue in any clean system. We further identify a strong-coupling phase with $C = 0$ that is nonetheless topologically nontrivial, characterized by a distinct Chern number $C^{(\mathrm{S})} \neq 0$ over the manifold of classical spin configurations. This phase is also disorder-driven, as $C^{(\mathrm{S})} = 0$ in the clean limit.
\end{abstract}

\maketitle

\section{Introduction}
\label{sec:intro}

Topologically nontrivial insulating phases at zero temperature are characterized by symmetry-protected topological invariants \cite{TKNN82,ASS83,Sim83,KM05,HK10,QZ11,CTSR16}. 
These invariants remain robust against weak symmetry-preserving perturbations as long as the bulk band gap does not close.
Chern insulators \cite{Hal88} in two dimensions, which exhibit the quantum anomalous Hall effect, provide a paradigmatic example.
They feature chiral conducting states along their one-dimensional boundaries and display a quantized Hall conductivity $\sigma = (e^{2}/ h) C$ determined by the integer-valued Chern number $C$.

Disorder (and similarly electron correlations \cite{Rac18}) introduces important challenges to the theory of topological insulators. 
In disordered systems, the notion of a band gap must be replaced by that of a {\em mobility gap}. 
First, electronic states may exist within this gap, but if they are Anderson localized \cite{And58}, they cannot support bulk electron transport. 
Localized and extended states are separated by mobility edges.
In two-dimensional systems, in particular, all eigenstates are localized for arbitrary finite disorder strength except at isolated critical energies where the localization length diverges \cite{AALR79,ATAF80}. 

Second, in clean systems the Chern number is typically understood as an invariant characterizing a bundle of Bloch states over the Brillouin zone. 
This relies on translational symmetry and therefore breaks down in disordered systems.
This difficulty was addressed in early work \cite{NTW85,TW84,AS85}, which demonstrated that the Hall conductance of a disordered system can be computed using twisted boundary conditions (TBCs) and remains quantized provided the mobility gap containing the chemical potential is finite and the many-electron ground state is non-degenerate. 
In the clean limit, this approach reduces to the TKNN invariant \cite{TKNN82}.


The TBC approach is closely related to real-space formulations of the Chern number \cite{LKZL23}, including those based on non-commutative geometry \cite{BvESB94,Pro11b,PHB10,Pro10} and on the Bott index \cite{LH11,HL10,Lor15}. 
In exceptional cases, invariance properties of the Chern number allow exact computation of phase diagrams directly in the thermodynamic limit \cite{PHB10}.
In most practical computations, however, TBCs are used for systems of finite size. 
In these cases, the convergence of real-space Chern numbers in the thermodynamic limit must be carefully checked numerically. 
The discussion in Ref.\ \cite{Pro11b} shows that this is not always straightforward.

A third issue -- and the focus of the present work -- is whether disorder can also play a {\em constructive} role, i.e., whether sufficiently strong disorder can induce a topologically nontrivial state. 
This question emerged in numerical studies \cite{LCJS09} of the BHZ model \cite{BHZ06} with uncorrelated onsite disorder, relevant for HgTe/CdTe quantum wells. 
In the clean limit, this system realizes a $\mathbb{Z}_{2}$ quantum spin Hall insulator.
Starting from the trivial phase and increasing the disorder strength $W$, numerical evidence indicated a transition to a topologically nontrivial state, inferred from the appearance of helical edge states. 
Because this phase occurs at Fermi energies where the clean system is metallic and has a non-inverted band structure, it was termed a disorder-induced topological Anderson insulator (TAI). 
The result was confirmed by several subsequent works \cite{GWA+09,JWSX09,YNIK11} and motivated searches for analogous phenomena in other systems \cite{GRRF10,BR11,LYR+20,LLX+22}. 
However, it was later shown \cite{Pro11a} that the ``TAI'' region in the disordered BHZ model is continuously connected to the quantum spin Hall phase once the mass parameter is included in the parameter space; it is therefore not a distinct topological phase. 

Here, we revisit the possibility of a constructive role of disorder, but for a different system: a spinful variant of a ``half-BHZ'' model with explicitly broken time-reversal symmetry—the spinful Qi-Wu-Zhang (QWZ) model \cite{QWZ06}, which hosts Chern-insulator phases similar to those of the Haldane model \cite{Hal88}. 
We furthermore consider a qualitatively different form of randomness, namely magnetic (spin) disorder. 
Specifically, the local electron spin $\ff s_{i\alpha}$ at each site $i$ and orbital $\alpha$ is coupled to classical unit-length spins $\ff S_{i\alpha}$ of random orientation, with the exchange coupling $J$ acting as the disorder strength.

We map out the topological phase diagram by computing the Chern number $C$ as a function of $J$ and the mass parameter $m$ using the TBC approach. 
The resulting phase structure is supported and refined by an independent method based on the Ishikawa--Matsuyama invariant \cite{IM86,Vol03}, expressed in terms of the disorder-averaged single-particle Green's function $\langle G \rangle$, in analogy with the treatment of interacting topological systems \cite{WZ12,Rac18}. 
After disorder averaging, $\langle G \rangle$ regains translational invariance and can be expressed in momentum space. 
To compute the Chern number, we employ the topological Hamiltonian (TH) \cite{WZ12}. 
We first adopt a local approximation for the disorder self-energy and subsequently include its momentum dependence. 
In addition, we consider the recently introduced S-space Chern number \cite{MFQ+24}, which topologically characterizes the eigenstate bundle over the manifold of classical spin configurations.

Our results reveal a highly intricate phase diagram in the $m$-$J$ parameter space, with multiple topologically trivial and nontrivial regions. 
Remarkably, we identify nontrivial phases in the strong-disorder regime that cannot be realized in the clean $J=0$ limit or in any other clean system.
These phases are characterized by a Chern number $C \ne 0$ generated either by zeros of the disorder-averaged Green’s function crossing the chemical potential or by a nonzero S-space Chern number $C^{\rm (S)} \ne 0$.

The paper is organized as follows:
The next section introduces the model and the type of disorder studied here. 
After a brief review of the twisted-boundary-condition approach and computational details in Secs.\ \ref{sec:tbc} and \ref{sec:comiss}, respectively, we discuss the resulting topological phase diagram in Sec.\ \ref{sec:res}. 
Section \ref{sec:green} introduces the Ishikawa--Matsuyama invariant and the topological Hamiltonian in the context of disordered systems. 
Results obtained within the local self-energy approximation are presented in Sec.\ \ref{sec:resth}, while nonlocal effects are discussed in Sec.\ \ref{sec:nonloc}. 
Section \ref{sec:space} introduces and applies the S-space Chern number. 
A summary and concluding discussion are given in Sec.\ \ref{sec:con}.

\section{Chern insulator with magnetic disorder}
\label{sec:mod}

We consider the spinful Qi-Wu-Zhang (QWZ) model \cite{QWZ06} on the two-dimensional square lattice as a prototypical Chern insulator with an additional disorder term. 
The Hamiltonian is given by $H=H_{0} + V$, where the translationally invariant (``clean'') part,
\be
H_{0} = \sum_{\ff k \alpha \alpha' \sigma} \epsilon_{\alpha\alpha'}(\ff k) c_{\ff k \alpha\sigma}^{\dagger}c_{\ff k\alpha'\sigma}
\: , 
\labeq{h0}
\ee
represents the QWZ model on a finite but large lattice with $L=L_{x} \times L_{y}$ sites ($L_{x}=L_{y}$) with 
periodic or twisted boundary conditions. 
Here, $\alpha = \text{A}, \text{B}$ is the orbital index, and $\sigma = \uparrow,\downarrow$ is the spin-projection index by which the model is trivially made spinful: $H_{0} = H_{0\uparrow} +H_{0\downarrow}$. 
The $\ff k$ sum runs over all allowed wave vectors of the first Brillouin zone (BZ), and $\epsilon_{\alpha\alpha'}(\ff k)$ are the elements of the $\ff k$-dependent $2 \times 2$ Bloch matrix,
\ba
  \ff \epsilon(\ff k)
  &=&
  \left[
  {m} + t \cos (k_{x}) + t \cos (k_{y}) 
  \right]
  \ff \tau_{z}
\nonumber \\
  &+&
  t \sin(k_{x}) \ff \tau_{x} + t \sin(k_{y}) \ff \tau_{y}
   \: .
\labeq{hk}
\ea
Here, $\tau_{x}, \tau_{y}, \tau_{z}$ denote the Pauli matrices. 
We set the hopping parameter to $t=1$ to fix the energy scale. 
The chemical potential is taken as $\mu=0$.

The band structure of the clean model is easily obtained as
$\epsilon_{\pm}(\ff k) = \pm [ t^{2} \sum_{r=x,y} \sin^{2}k_{r} + ({m}+ t \sum_{r=x,y} \cos k_{r})^{2} ]^{1/2}$.
At critical values $m_{\rm c}=-2t,0,2t$ of the mass parameter $m$, the band gap 
$\Delta = \mbox{min} \{ \epsilon_{+}(\ff k) - \epsilon_{-}(\ff k) \, | \, \ff k \in \mbox{BZ}  \}$
closes at critical high-symmetry points in the BZ, at $\ff k_{\rm c} = (0,0)$, at $\ff k_{\rm c}=(0,1),(1,0)$, or at $\ff k_{\rm c}= (1,1)$, respectively. 
For $m=m_{\rm c}$, the low-energy effective theory is given by a linear Dirac model with $\ff k$-independent mass term.
In the gapped phases, for $-2t<m<0$ and for $0<m<2t$, the Chern number is $C=-2$ and $C=+2$, respectively, while $C=0$ for $|m|>2t$.
The spinless QWZ model is invariant under a unitarily represented particle-hole transformation that squares to the identity. It is not invariant under time reversal.
Thus, it belongs to symmetry class D in the Altland-Zirnbauer (AZ) classification scheme \cite{AZ97,RSFL10} with $C \in \mathbb{Z}$.
The reason that $C \in 2 \mathbb{Z}$ rather than $C \in \mathbb{Z}$ is simply that the spinful QWZ model is obtained by trivial additive doubling of the original spinless model.

The disorder term $V$ is most conveniently specified in the real-space representation of the model where a site-orbital-spin one-particle basis $\{ | i, \alpha, \sigma \rangle \}$ is used.
The clean part in this representation, 
$H_{0} = \sum_{ii'} \sum_{\alpha\alpha'} \sum_{\sigma} t_{i \alpha, i' \alpha'} c^{\dagger}_{i \alpha\sigma} c_{i' \alpha' \sigma}$,  is obtained via Fourier transformation
$c^{\dagger}_{\ff k \alpha \sigma} = \frac{1}{\sqrt{L}} \sum_{i} e^{i \ff k \ff R_{i}} c^{\dagger}_{\ff R_{i} \alpha \sigma}$, 
where $\ff R_{i}$ denote the lattice vectors.
The disorder term has the form of a local exchange coupling, 
\be
V = J \sum_{i=1}^{L} \sum_{\alpha=A,B} \ff s_{i\alpha} \cdot \ff S_{i\alpha}
\: , 
\labeq{dis}
\ee
where $\ff s_{i\alpha} = \frac12 \sum_{\sigma, \sigma'}c^{\dagger}_{i\alpha\sigma} \ff \tau_{\sigma \sigma'} c_{i\alpha \sigma'}$ is the electron spin at site $i$ and orbital $\alpha$, where $\ff \tau = (\tau_{x}, \tau_{y}, \tau_{z})$, and where $\ff S_{i\alpha}$ is a ``classical'' spin, i.e., a local magnetic moment of fixed length $|\ff S_{i\alpha}|=1$. 
The $\ff S_{i\alpha}$ for different $i,\alpha$ are uncorrelated random variables, each uniformly distributed over a 2-sphere $\mathbb{S}^{2}$, i.e., \refeq{dis} describes local, uncorrelated, directional disorder of classical spins. 
The disorder strength is given by the strength of the exchange coupling $J$. 
The clean limit is given by $J=0$.
The set of all configurations $S \equiv (\ff S_{1A}, \ff S_{1B}, ..., \ff S_{LA}, \ff S_{LB})$ of the classical spins (with $|\ff S_{i\alpha}| = 1$) forms the $4L$-dimensional S-space manifold $\mathbb{S} = \mathbb{S}^{2} \times \cdots \times \mathbb{S}^{2}$.
The Hamiltonian $H=H(S)$, and thus its eigenenergies and eigenstates depend on the spin configuration $S$.

\section{Twisted boundary conditions}
\label{sec:tbc}

For a generic random spin configuration $S \in \mathbb{S}$, the system lacks translational symmetries and, hence, the eigenstates of $H=H(S)$ can no longer be characterized by wave vectors $\ff k \in {\rm BZ}$. 
In a first step we replace periodic boundary conditions by twisted boundary conditions (TBC) \cite{TKNN82,LKZL23,BvESB94}.
The hopping amplitudes $t_{x}$ linking edge sites $\ff R=(1, R_{y})$ and $\ff R=(L_{x} ,R_{y})$ along the $x$ direction are replaced as $t_{x} \to t_{x} e^{i \theta_{x}}$, and analogous for the $y$ direction, $t_{y} \to t_{y} e^{i \theta_{y}}$. 
With this, the model is defined on a 2-torus $\mathbb{T}^{2} \cong \ca T = \{(\theta_{x},\theta_{y}) \, | \, \theta_{x},\theta_{y} \in [0,2\pi] \}$ spanned by the twist angles, and we have $H = H(\theta_{x},\theta_{y}) = H(\theta_{x}+2\pi,\theta_{y}) = H(\theta_{x},\theta_{y}+2\pi)$.
Diagonalization yields an orthonormal basis of single-particle eigenstates $| \psi_{n}(\ff \theta) \rangle$ for each pair of twist angles $\ff \theta=(\theta_{x},\theta_{y})$.

We consider the spinful model with spin projection $\sigma = \,\uparrow, \downarrow$ and with two orbitals $\alpha=\mbox{A, B}$ at half-filling, such that the energy spectrum consists of $2L$ occupied and of $2L$ unoccupied eigenstates. 
We assume that, for each point in $\ca T$, there is no gap closure due to a delocalized state, such that the projector onto the occupied states, defined as $P_{\ff \theta} = P(\theta_{x},\theta_{y}) = \sum_{n}^{\rm occ.} | \psi_{n}(\ff \theta) \rangle \langle \psi_{n}(\ff \theta) |$, is a smooth function on $\ca T$.
With this we get a well-defined Chern number $C(S) \in 2 \mathbb{Z}$ for a given spin configuration $S$ as
\be
C(S) = \frac{-1}{2\pi i} 
\int_{0}^{2\pi} \! \! \int_{0}^{2\pi} d\theta_{x} d\theta_{y} \, 
\mbox{tr} \Big(
P_{\ff \theta} \, [\partial_{\theta_{x}} P_{\ff \theta} , \partial_{\theta_{y}} P_{\ff \theta}]
\Big) \: .
\labeq{chern}
\ee
In the clean limit the Chern number, given by \refeq{chern} and obtained via TBC, is the same as the Chern number that is obtained with the BZ as the base manifold \cite{LKZL23}.

Operationally, at finite disorder strength, we draw independent configurations $S$ from the local uncorrelated distribution on $\mathbb{S}$. 
For each $S$, the Hamiltonian $H(S)=H_{0} + V(S)$ is diagonalized to construct the eigenstates $| \psi_{n}(\theta_{x},\theta_{y}) \rangle$ at a random pair $\theta= (\theta_{x},\theta_{y})$ of twist angles. 
The occupied eigenstates and their derivatives w.r.t.\ $\theta_{x}$ and $\theta_{y}$ are used to construct the projector $P_{\theta}$. 
Furthermore, we profit from the fact that there is no significant dependence on the twist angles \cite{KWKH19}, such that the integration in \refeq{chern} is trivially achieved by multiplying the integrand by the 2-torus surface area $4\pi^{2}$. 
Calculations of $C(S)$ are repeated for different $S$ to obtain the average Chern number $C$ and its standard deviation.

\section{Phase diagram and computational details}
\label{sec:comiss}

\begin{figure}[t] 
\centering
\includegraphics[width = 1.0\linewidth]{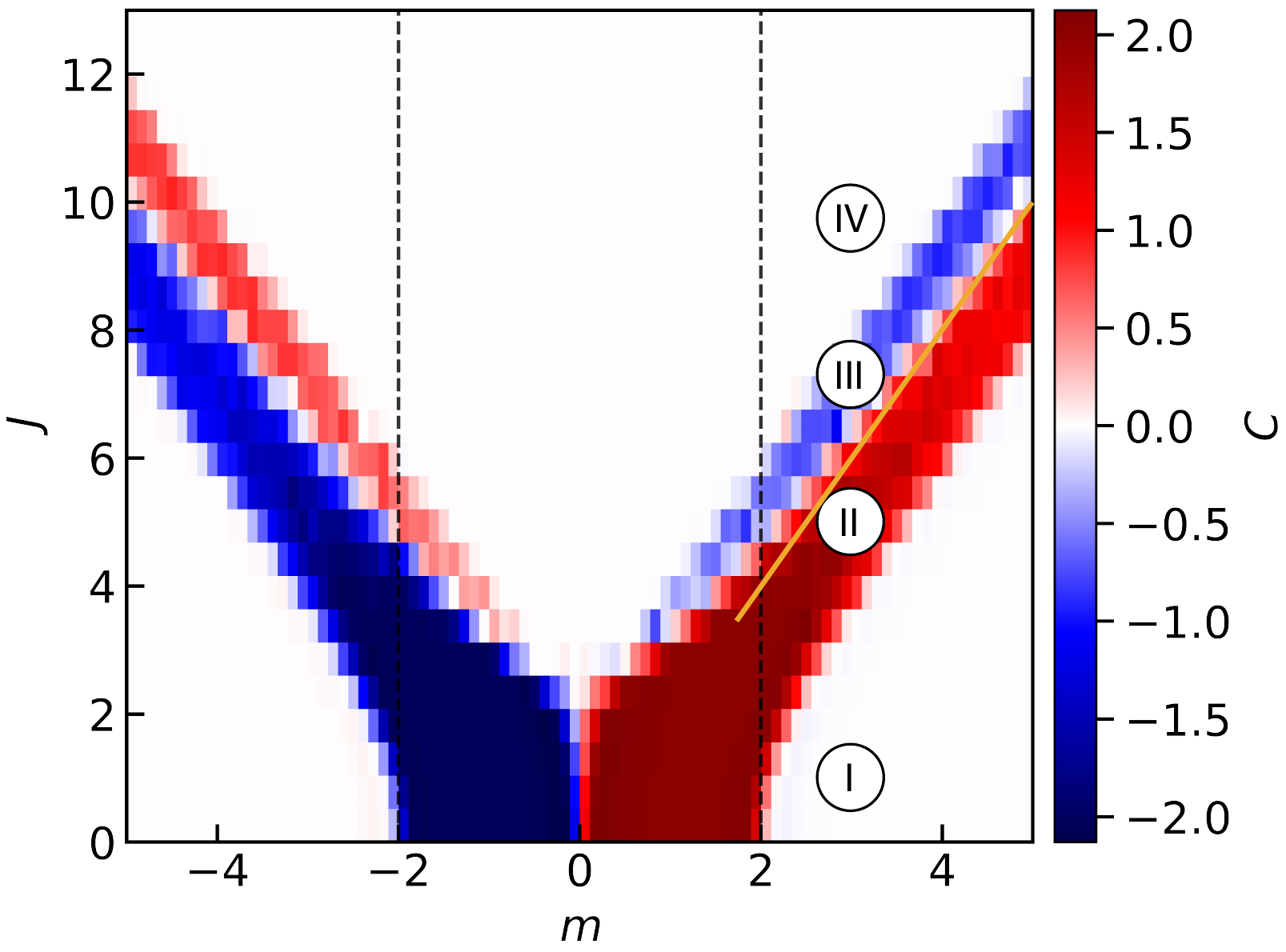}
\caption{
Disorder-averaged Chern number (color code) as obtained from TBC approach as function of the mass parameter $m$ and disorder strength $J$. 
Results obtained for a system with $L_{x} \times L_{y} =10 \times 10$ sites and $N_{\rm conf.} = 50$ disorder configurations. 
Chemical potential $\mu=0$.
See text for discussion of phases I, II, III, IV and the full yellow line.
}
\label{fig:pd1}
\end{figure}

\begin{figure}[t] 
\centering
\includegraphics[width = 0.9\linewidth]{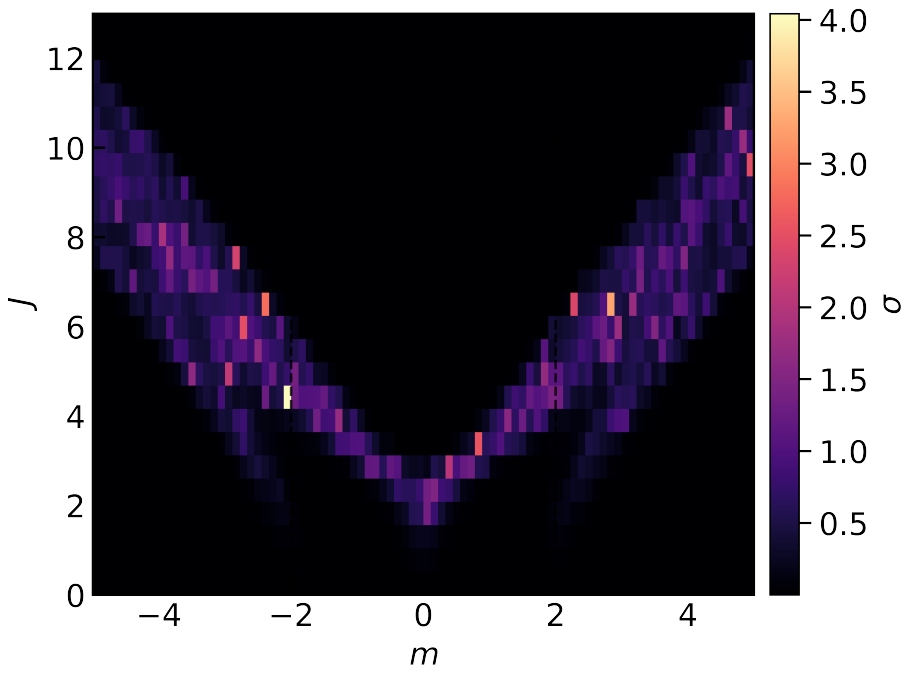} 
\includegraphics[width = 0.9\linewidth]{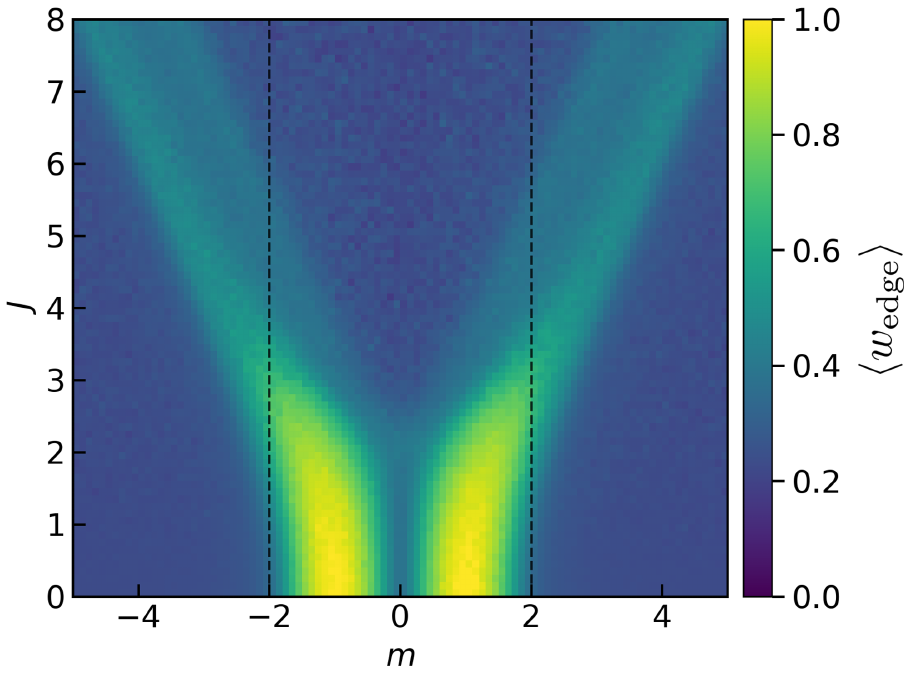}
\caption{
{\em Upper panel:}
Standard deviation of the Chern number $C$ corresponding to the average shown in Fig.\ \ref{fig:pd1}.
{\em Lower panel:}
Disorder-averaged ($N_{\rm conf.}=300$) weight $\langle w_{\rm edge} \rangle$ of the zero-energy (spin-degenerate) chiral eigenstate at the edge sites of same system as in Fig.\ \ref{fig:pd1} but with {\em open} boundaries, as function of $m$ and $J$.
The total weight summed over {\em all} sites is normalized to unity.
}
\label{fig:pd2}
\end{figure}

We have performed numerical calculations for systems of finite size using twisted boundary conditions.
The resulting disorder-averaged Chern number $C$, as obtained from \refeq{chern} for a sample of $N_{\rm conf}=50$ random spin configurations $S \in \mathbb{S}$, turns out as quantized with values $C=-2,0,2$ for parameters deep in a gapped phase, as expected.
This is seen in Fig.\ \ref{fig:pd1} in the clean limit ($J=0$) and in the dark blue ($C=-2$) and dark red ($C=2$) colored phases up to moderate disorder strength $J\approx 2$.
Contrary, at phase boundaries and also in parametric vicinity to the boundaries, non-integer values for $C$ are obtained numerically.
For example, at strong $J \gtrsim 4$ and $m>0$ within the (red colored) phase II, we find that $C$ deviates from the integer value $C=2$.
 
Non-integer values of $C$ are due to non-interger values of $C(S)$, i.e., of the Chern numbers for individual spin configurations [see \refeq{chern}] and must be interpreted as finite-size artifacts. 
Quantization of $C(S)$ requires that states at $\mu = 0$ be localized, making them insensitive to changes of the twist angles and preventing them from contributing to the integral in \refeq{chern}.
Conversely, the presence of a delocalized state at $\mu=0$ is necessary for a topological phase transition \cite{Pro11b}.
For a system of finite size, states with a localization length $\xi_{\rm loc} \sim L_{x}, L_{y}$, however, do contribute to $C(S)$ and thus destroy the quantization. 
This is a well-known finite-size artifact, which has been discussed in Refs.\ \cite{Pro11a,LKZL23}, for example. 

The calculations underlying Fig.\ \ref{fig:pd2} have been performed for $L=L_{x} \times L_{y} = 10^{2}=100$, i.e., for 400 spin orbitals. 
The results for the disorder-averaged Chern number $C$ shown in the figure are well converged with respect to the sample size ($N_{\rm conf.} = 50$). 
Note that less configurations are required with increasing $L$, since $C$ is self-averaging. 
Averaging over different spin configurations smoothens the results but does not reinforce quantization.
It is important to stress that the size dependence turns out as crucial, and with $L=\ca O(10^{2})$ one is still far from converged and integer values for $C$. 

Quantized values $C=-2,0,2$ expected in the thermodynamical limit $L\to \infty$.
The deviations from these integers seen in Fig.\ \ref{fig:pd2} are consistent with the computed standard deviation $\sigma$ of the Chern number, which is shown in Fig.\ \ref{fig:pd2} (upper panel). 
Deep in the $C=\pm2$ and $C=0$ phases, $\sigma$ vanishes, while for strong $J$ and in particular close to the phase boundaries the deviation is large, $\sigma \approx 1$.

For a crosscheck, additional calculations were performed for the same system but with open boundaries. 
In this case, due to the bulk-boundary correspondence, we expect a dispersive chiral edge mode to cross the gap. 
Figure \ref{fig:pd2} (lower panel) shows the results obtained for the total weight, $w_{\rm edge}$, of the (spin-degenerate) eigenstate with closest energy to $\mu=0$ on the edge sites. 
For weak $J$ and for $m$ deep in the $C=\pm 2$ phases, the edge mode is strongly localized on the edge sites, with almost negligible weight spreading into the bulk, i.e., $w_{\rm edge} \lesssim 1$.
For strong $J$ and in the topologically nontrivial phases, the total weight on the edge sites decreases though it remains significantly higher than $w_{\rm edge}$ in the $C=0$ phase. 
A vanishing $w_{\rm edge}$ is obviously possible only in the limit of infinite lattices ($L\to \infty$). 

One important difference of magnetic disorder as compared to potential disorder with, e.g., the on-site energies $\varepsilon_{i} \in [-W/2 , W/2]$ distributed homogeneously within an energy range $W$, is that energy eigenstates are generically less localized. 
This makes a full finite-size scaling computationally highly demanding.
The stronger delocalization in the case of magnetic disorder is seen as follows: 
Addressing the strong-disorder (strong $J$) regime by expanding around the $t=0$ atomic limit, there is an overlap $\propto \cos(\vartheta/2)$ between two atomic states at neighboring sites $i$, $j$, which is relevant at second order in $t$, where $\vartheta$ is the angle enclosed by the respective classical spins. 
This is generically finite, such that the overlap is of the order $\ca O(1)$.
On the other hand, the generic overlap between states localized at neighboring sites is exponentially small in the case of potential disorder for $W\to \infty$.

\section{Discussion of the topological phase diagram}
\label{sec:res}

The unitary gauge transformation $c_{i\alpha\sigma} \mapsto (-1)^{i} c_{i\alpha\sigma}$, where $(-1)^{i}$ is the staggered sign factor on the bipartite lattice, maps $H=H(m,S)$, given a mass parameter $m$ and a spin configuration $S$, to $-H(-m,S)$. 
The Chern number of the transformed system, and therefore of the system with the Hamiltonian $+H(-m,S)$, changes sign, i.e., $C(-m,S) = - C(m,S)$.
Hence, the topological phase diagram is antisymmetric with respect to the $m=0$ axis.
When discussing the phase diagram in Fig.\ \ref{fig:pd1}, we exploit this antisymmetry and focus primarily on the range $m \ge 0$. 

The topologically trivial and nontrivial phases with Chern numbers $C=0$ and $C=+2$ in the clean limit (phases I and II in Fig.\ \ref{fig:pd1}) continuously extend to finite $J$. 
For weak disorder, $J \lesssim 2$, the boundary between I and II is weakly $J$ dependent, and phase II only slightly extends to mass parameters $m>2$.

For fixed mass parameter in the range $0< m < 2$ and increasing $J$, phase II with $C=2$ (in red) continuously evolves from the $J=0$ phase until, at a critical disorder strength $J_{\rm c}(m)$ (for small $m$ within the range $2.5 \lesssim J_{\rm c} \lesssim 3.5$), there is a transition to a different nontrivial phase III with $C=-2$ (at $m>0$, blue color).
This transition from phase II to III will be discussed more precisely below. 
At still stronger $J$, we find a second transition at $J_{\rm c2}(m) > J_{\rm c}(m)$, and the system ends up in a trivial phase with $C=0$ (phase IV).

For $m > 2$, the system is in the trivial $C=0$ phase I at $J=0$. 
Keeping $m>2$ fixed and increasing $J$, we find three phase transitions with $m$-dependent boundaries:
from phase I to phase II at another critical coupling $J_{\rm c1}(m)$, from phase II to phase III at $J_{\rm c}(m)$, and finally from phase III to phase IV at $J_{\rm c2}(m)$.
These critical couplings $J_{\rm c1} < J_{\rm c} <J_{\rm c2}$ are strongly $m$ dependent and become linear-in-$m$ functions for large $m$. 
The situation for $m \to 0$, on the other hand, is less clear due to the lack of data for large system sizes (see also Sec.\ \ref{sec:green}).

For strong disorder, $3 \lesssim J < \infty$, there is a trivial phase around $m=0$, and its $m$ extension grows as $J$ increases.
This $C=0$ phase IV separates the two $C=-2$ ($m<0$) and $C=2$ ($m>0$) phases, which are connected to the corresponding weak-$J$ phases. 
However, it furthermore separates two new phases with $C=+2$ and $C=-2$ for $m<0$ and $m>0$, respectively.
These phases are only found at strong $J$.

For phase II, this is different:
While for fixed $m>2$, the clean system is topologically trivial ($C=0$), the nontrivial phase II $(C=2$) emerges at finite and sufficiently strong $J>J_{\rm c1}(m)$, see the dashed black lines in Fig.\ \ref{fig:pd1}.
This situation is reminiscent of a long-standing discussion of a novel nontrivial ``topological Anderson insulator'' phase found \cite{LCJS09} in the BHZ model \cite{BHZ06} with local potential disorder, which is disorder-induced, i.e., topologically trivial when disorder is switched off. 
However, it finally turned out \cite{Pro11a} that the phase is continuously connected with a corresponding clean-limit phase and, therefore, actually does not represent a new phase.
Phase III, on the other hand, cannot be connected continuously by varying $m$, $J$, or the chemical potential $\mu$ to the $C=-2$ phase at weak-$J$ seen in the negative $m$ range (blue color). 
We emphasize that, at this point, one cannot exclude that the two phases could finally turn out as continuously connected via extra dimensions in parameter space and that, on the other hand, equal Chern numbers $C=-2$ does not imply that such a continuous connection must exist.
However, we will come back to this issue in Sec.\ \ref{sec:green} below.

\begin{figure}[t] 
\centering
\includegraphics[width = 0.8\linewidth]{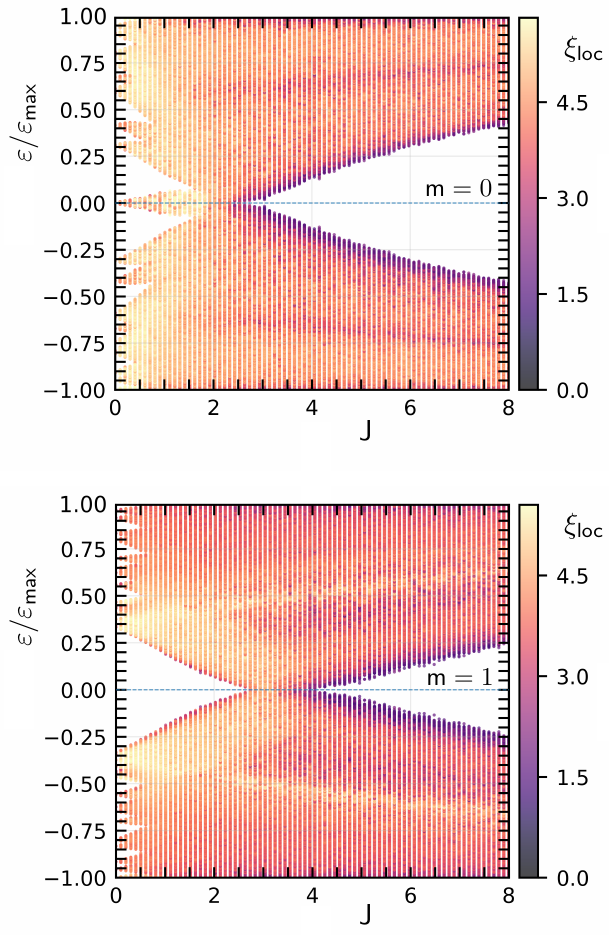}
\caption{
Eigenenergies $\varepsilon$ and localization length $\xi_{\rm loc}$ (color code) of corresponding eigenstates, as function of the disorder strength $J$.
{\em Upper panel: } $m=0$. 
{\em Lower panel: } $m=1$. 
$\xi_{\rm loc}$ as obtained from the inverse participation ratio, see text.
The energy spectrum is normalized to the maximum eigenenergy $\varepsilon_{\rm max}$ for each $J$ (for $J=0$: $\varepsilon_{\rm max} = 2$ at $m=0$ and $\varepsilon_{\rm max} = 3$ at $m=1$).
System size: $L_{x} \times L_{y} =10 \times 10$ sites. 
$N_{\rm config}=10$.
Chemical potential $\mu=0$.
}
\label{fig:spec}
\end{figure}

Generally, any transition between phases with different Chern numbers requires the presence of a delocalized eigenstate at $\mu$, see, e.g., Ref.\ \cite{XP13} and references therein.
The energy of such a state with divergent localization length marks a boundary of a mobility gap. 
To quantify the degree of localization, we have computed for each eigenstate $| \psi \rangle$ (with $\langle \psi | \psi \rangle=1$) the inverse participation ratio $\rho = \sum_{i\alpha\sigma} |\langle i, \alpha, \sigma | \psi \rangle |^{4}$. 
This can be used \cite{FM92} to get an estimate for the localization length $\xi_{\rm loc}$ via $\xi_{\rm loc} = 1/\sqrt{2\pi \rho}$ (if $\xi_{\rm loc} \ll L_{x}, L_{y}$).

A generic example is given with Fig.\ \ref{fig:spec}, which provides an overview of the spectrum of eigenenergies $\varepsilon$ and the corresponding localization lengths $\xi_{\rm loc}$ at $m=0$ (upper panel) and $m=1$ (lower panel) and as a function of $J$. 
Note that there is a particle-hole symmetry of the model for arbitrary $m$, which implies that the spectrum is symmetric to $\varepsilon=0$.

In the clean $J=0$ limit, the system is in a semi-metallic state at $m=0$ with a gap closure at $\ff k = (0,\pi)$. 
At finite disorder strength up to $J \lesssim 2$, the states at $\varepsilon=\mu=0$ remain delocalized, i.e., the respective localization length remains close to the linear size of the system. 
For $J\approx 2.5$, a spectral gap around $\varepsilon=0$ opens and widens with inreasing $J$.
Only shortly before the gap reopens, the localization length strongly drops to $\xi_{\rm loc} \approx 1$.

For $m=1$ (see lower panel of the figures), the band gap $\Delta = 2$ ($\Delta / \varepsilon_{\rm max} = 2/3$) is at a maximum in the clean limit. 
The spectral gap shrinks with increasing $J$.
For coupling strengths greater than $J\approx 2.7$, the system exhibits states at $\varepsilon=\mu=0$ but with a localization length $\xi_{\rm loc}$ smaller than the linear system size. 
At $J \approx 3$, however, $\xi_{\rm loc}$ reaches a maximum with $\xi_{\rm loc} \approx L_{x}=L_{y}$.
Interpreting this maximum as the appearance of a delocalized state at $\mu=0$, is consistent with the critical disorder strength  $J_{\rm c} \approx 3$ at $m=1$ that can be read off from the phase diagram in Fig.\ \ref{fig:pd1}. 

With increasing $m$, this critical coupling $J_{\rm c} = J_{\rm c}(m)$, separating phases II and III, becomes a linear function of $m$. 
This observation and the very existence of a delocalized state at $\mu=0$ can easily be understood analytically:
We focus on mass parameters $m\ge 0$, for simplicity, and arrange the Hamiltonian $H=H_{0}+V$, see Eqs.\ (\ref{eq:h0}) and (\ref{eq:dis}), as $H=H_{\rm local} + H_{\rm nonlocal}$, where the local part consists of the mass and the disorder terms:
\be
H_{\rm local} 
= 
\sum_{i\alpha \alpha' \sigma \sigma'} 
\Big[
m 
\delta_{\sigma\sigma'} \tau_{z,\alpha\alpha'}
+
\frac{J}{2} \delta_{\alpha\alpha'}
\ff \tau_{\sigma \sigma'}\ff S_{i\alpha} 
\Big]
c^{\dagger}_{i\alpha\sigma} c_{i\alpha'\sigma'} 
\: .
\labeq{hloc}
\ee
The remaining nonlocal part $H_{\rm nonlocal}$ is just the (spinful and clean) QWZ model at $m=0$.
Diagonalization of the atomic $4 \times 4$ Hamilton matrix in \refeq{hloc} at a site $i$, yields two eigenvalues $\varepsilon_{\pm,{\rm high}} = \pm (m+J/2)$, which diverge for $m,J \to \infty$, and two eigenvalues $\varepsilon_{\pm,{\rm low}} = \pm (m-J/2)$, which stay finite if $\delta \equiv 2m-J$ is finite. 
In the $m,J \to \infty$ limit with $\delta=\mbox{const}$, the low-energy eigenstates are given by 
$| \varepsilon_{+,{\rm low}} \rangle = | {\rm A} \rangle \otimes |  - \ff S_{i\rm A} \rangle \equiv | -\ff S_{i\rm A} \rangle$, i.e., an $\alpha=\text{A}$ orbital with the electron spin aligned antiparallel to the classical spin $\ff S_{i\rm A}$ at that site, and analogously $| \varepsilon_{-,{\rm low}} \rangle = | + \ff S_{i\rm B} \rangle$. 
With the projector 
\be
P_{\rm low} = \sum_{i} 
\Big(
| \varepsilon_{+,{\rm low}} \rangle
\langle \varepsilon_{+,{\rm low}} |
+
| \varepsilon_{-,{\rm low}} \rangle
\langle \varepsilon_{-,{\rm low}} |
\Big)
\ee
we can project out the high-energy states, $H_{\rm local} \mapsto P_{\rm low} H_{\rm local} P_{\rm low}$.
This leaves us, to leading order, with an effective Hamiltonian given by
\ba
H_{\rm eff}
&=&
\frac12 \delta
\sum_{i}
\Big(
| - \ff S_{i\rm A} \rangle \langle - \ff S_{i\rm A} |
-
| +\ff S_{i\rm B} \rangle \langle +\ff S_{i\rm B} |
\Big)
\nonumber \\
&+&
H_{\rm QWZ, m=0}
\: .
\labeq{heff}
\ea
This effective model applies to the $m,J \to \infty$ regime of the phase diagram, where $\delta = 2m - J =\ca O(t)$ is small. 
It comprises the $m=0$ QWZ model with an additional local disordered part proportional to $\delta$.
Importantly, for $\delta=0$, only the QWZ term remains, and thus all eigenstates are fully delocalized.
Furthermore, since only the QWZ model at $m=0$ contributes, the system is critical. 
We conclude that, in the limit $m,J \to \infty$, the full model is gapless on the line $J=2m$. 
Hence, this strictly implies the existence of an extended state at the chemical potential. 
Asymptotically, for large $m$, the $J=2m$ line in fact perfectly corresponds to the phase boundary between the $C=2$ and the $C=-2$ topological phase, see the full yellow line in Fig.\ \ref{fig:pd1}. 

The effective model can be also written in the form
\ba
H_{\rm eff}
&=&
\frac12 \delta
\sum_{i \alpha \alpha' \sigma \sigma'}
\frac12 
\Big(
\delta_{\sigma \sigma'} \tau^{(z)}_{\alpha\alpha'} - \ff S_{i\alpha} \cdot \ff \tau_{\sigma\sigma'} \delta_{\alpha\alpha'}
\Big)
c^{\dagger}_{i\alpha\sigma} c_{i\alpha'\sigma'} 
\nonumber \\
&+&
H_{\rm QWZ, m=0}
\: .
\labeq{heff2}
\ea
For small $\delta \ne 0$, this is simply the original Hamiltonian but with renormalized parameters $m \mapsto m' = \delta / 4$ and $J \mapsto J' = - \delta / 2$. 
This shows that in the limit $m, J \to \infty$ with small $\delta = 2m - J > 0$ (just below the yellow line in the figure) we have $C=+2$, as in the original system with $m=\delta / 4$ and $J=-\delta / 2$. 
Here, we exploit the fact that the topological phase diagram is symmetric under the parameter transformation $J \to -J$, since this is compensated by a transformation of the spin configuration $S \mapsto -S$ and thus leaves the Chern number invariant. 
Analogously, we have $C=-2$ for small $\delta = 2m - J < 0$. 
The Chern numbers $C=\pm 2$ are consistent with our expectation that the numerical data for $C$ at strong $m$ and $J$ should converge to $C = \pm 2$ in the limit $L\to \infty$. 
Importantly, the mapping of the model for $m, J \to \infty$ onto the original model with $m', J'$ only holds for small $\delta$, since the neglected high-energy terms increasingly affect the delicate balance between strong disorder $J$, enforcing singly occupied (half-filled) orbitals $n_{i \alpha} = \sum_{\sigma=\uparrow,\downarrow} \langle c^{\dagger}_{i\alpha\sigma} c_{i\alpha\sigma} \rangle \to 1$, and a strong mass parameter, enforcing saturated orbital polarization $p = \frac12 (n_{i \rm B} - n_{i\rm A}) \to \pm 1$ with full or empty orbitals.

\section{Chern number from disorder averaged Green's function}
\label{sec:green}

We compare the phase diagram Fig.\ \ref{fig:pd1}, as obtained by using TBC, with the results of an alternative ``topological Hamiltonian'' (TH) approach. 
Here, we consider the frequency-dependent single-particle Green's function $\ff G^{(S)}(\omega)$ with elements $G^{(S)}_{i\alpha\sigma, i'\alpha'\sigma'}(\omega)$ for a given spin configuration $S \in \mathbb{S}$.
These are obtained by matrix inversion for a frequency $\omega \in \mathbb{C} \setminus \mathbb{R}$, i.e., $\ff G^{(S)}(\omega) = (\omega - \ff t)^{-1}$, from the hopping matrix $\ff t$ that represents the Hamiltonian $H=H_{0}+V$, see Eqs.\ (\ref{eq:h0}) and (\ref{eq:dis}), in the basis $\{ | i, \alpha, \sigma \rangle \}$:
\be
H = \sum_{ii'} \sum_{\alpha\alpha'\sigma\sigma'} 
t_{i\alpha\sigma, i'\alpha'\sigma'}
c^{\dagger}_{i\alpha\sigma}
c_{i'\alpha'\sigma'}
\; .
\ee
Here $\ff t = \ff t_{0} + \ff V$, and the elements of $\ff t_{0}$ are obtained by Fourier transformation from the $\ff k$-dependent Bloch-matrix elements $\epsilon_{\alpha\alpha'}(\ff k)$, see \refeq{hk}, while
$
V_{i\alpha\sigma, i'\alpha'\sigma'}
=
\frac12 J \delta_{ii'} \delta_{\alpha\alpha'}
\ff \tau_{\sigma \sigma'} \cdot \ff S_{i\alpha}
$.

The disorder-averaged Green's function $\ff G(\omega) \equiv \langle \ff G^{(S)}(\omega) \rangle$ is invariant under lattice translations. 
Therefore, after Fourier transformation to reciprocal space, it is diagonal with respect to the wave vector $\ff k$, and its elements are of the form $G_{\alpha\sigma, \alpha'\sigma'}(\ff k, \omega)$. 
Hence, $4 \times 4$ matrix inversion for each $\omega$ and $\ff k$ yields the disorder self-energy
\be
  \ff \Sigma(\ff k, \omega) = \ff G_{0}(\ff k, \omega)^{-1} - \ff G(\ff k, \omega)^{-1}
  \:, 
\labeq{sigma}
\ee
where $\ff G_{0}(\ff k, \omega)^{-1} = \omega + \mu - \ff \epsilon(\ff k)$ is the Green's function for $\ff V=0$.

Disorder averaged Green's functions and disorder self-energies are usually utilized in the context mean-field approximations, like the coherent-potential approximation, and cluster extensions, as discussed, e.g., in Refs.\ \cite{PB07,TKM+25} in the context of disordered and {\em correlated} electron systems. 
Here, we employ the Green's function to construct a homotopy invariant \cite{IM86,Vol03,Gur11}
\ba
N_{2} 
&=&
\frac{1}{24 \pi^{2}}
\iiint d(i\omega) dk_{x} dk_{y}
\nonumber \\
&\times&
\sum_{\mu\nu\rho}
\mbox{tr}
\left(
\varepsilon^{\mu\nu\rho} 
\ff G \partial_{\mu} \ff G^{-1}
\ff G \partial_{\nu} \ff G^{-1}
\ff G \partial_{\rho} \ff G^{-1}
\right)
\: , 
\labeq{im}
\ea
where $\mu, \nu, \rho$ run over $i \omega, k_{x}, k_{y}$ (with $i \omega \in i \mathbb{R}$).
In Ref.\ \onlinecite{WZ12} it is demonstrated that the inconvenient integration along the imaginary frequency axis can be avoided and that the Ishikawa-Matsuyama invariant \refeq{im} of a two-dimensional gapped interacting electron system can be computed equivalently from the Chern number $C$ of the much simpler fictive non-interacting ``topological Hamiltonian''. 
The argumentation in Ref.\ \onlinecite{WZ12} is based on very general analytical and causal properties of single-particle Green's functions and on the assumption that the Green's function of the interacting system is nonsingular, i.e., without a pole of the Green's function or of the self-energy at $\omega=0$. 
With the same assumption, we can transfer the argument to a disordered system. 
This means that the homotopy invariant $N_{2} \in \mathbb{Z}$ of the disordered system is equivalent to the Chern number $C$ obtained via the topological Hamiltonian
\be
H_{\rm top} 
= 
H_{0} 
+
\sum_{ii'} \sum_{\alpha\alpha'\sigma\sigma'} 
\Sigma_{i\alpha\sigma, i'\alpha'\sigma'}(\omega=i \eta)
c^{\dagger}_{i\alpha\sigma}
c_{i'\alpha'\sigma'}
\; ,
\labeq{toph}
\ee
from the disorder self-energy. Here, $\eta \searrow 0$ is a positive infinitesimal, and only the retarded zero-frequency self-energy matrix $\ff \Sigma \equiv \ff \Sigma (\omega + i \eta)|_{\omega=0}$ is required.

The absence of self-energy poles at $\omega=0$ is guaranteed, if the retarded disorder self-energy, \refeq{sigma}, is Hermitian.
In particular, we must have 
\be
  - \frac{1}{\pi}  \mbox{Im} \, \Sigma_{\alpha\sigma,\alpha\sigma}(\ff k, \omega + i \eta) \big|_{\omega=0} \stackrel{!}{=} 0 
  \: 
\labeq{imsig}  
\ee
for the diagonal elements.
This condition will be met if there is a finite spectral gap around $\omega=0$.
If the gap is a mobility gap only, the presence of localized states at $\omega=0$ will lead to a non-Hermitian contribution to the disorder self-energy.
However, since localized states cannot cause a change of the Chern number, 
we can ignore this part and only employ the Hermitian part of the self-energy in the topological Hamiltonian, \refeq{toph}.
In fact, this has been done in various previous approaches to systems with local on-site disorder, in particular using weak-coupling approaches such as the self-consistent Born approximation, see, e.g., Refs.\ \cite{GWA+09,YNIK11}.

\section{Topological phase diagram as obtained from the TH approach}
\label{sec:resth}

In a first step, we assume that it is sufficient to keep the $\ff k$-averaged, $\omega=0$ retarded self-energy 
\be
\ff \Sigma \equiv \frac{1}{L} \sum_{\ff k} \ff \Sigma(\ff k, \omega + i \eta)|_{\omega=0}
\labeq{kindep}
\ee
only.
We refer to this approach as the local approximation. 
The local approximation simplifies the computation of the Chern number considerably, since $\ff \Sigma$ is diagonal with respect to the orbital and the spin indices. 
Its elements have the form 
\be
\Sigma_{\alpha\sigma, \alpha'\sigma'} \equiv  \Sigma \, \delta_{\sigma\sigma'} \tau_{z,\alpha\alpha'}
\labeq{sform}
\ee
with a scalar quantity $\Sigma$.
This means that the disorder self-energy leads to a renormalization of the mass parameter only: 
\be
m \to m^{\ast} = m + \Sigma
\: .
\labeq{renor}
\ee

\begin{figure}[t] 
\centering
\includegraphics[width = 0.95\linewidth]{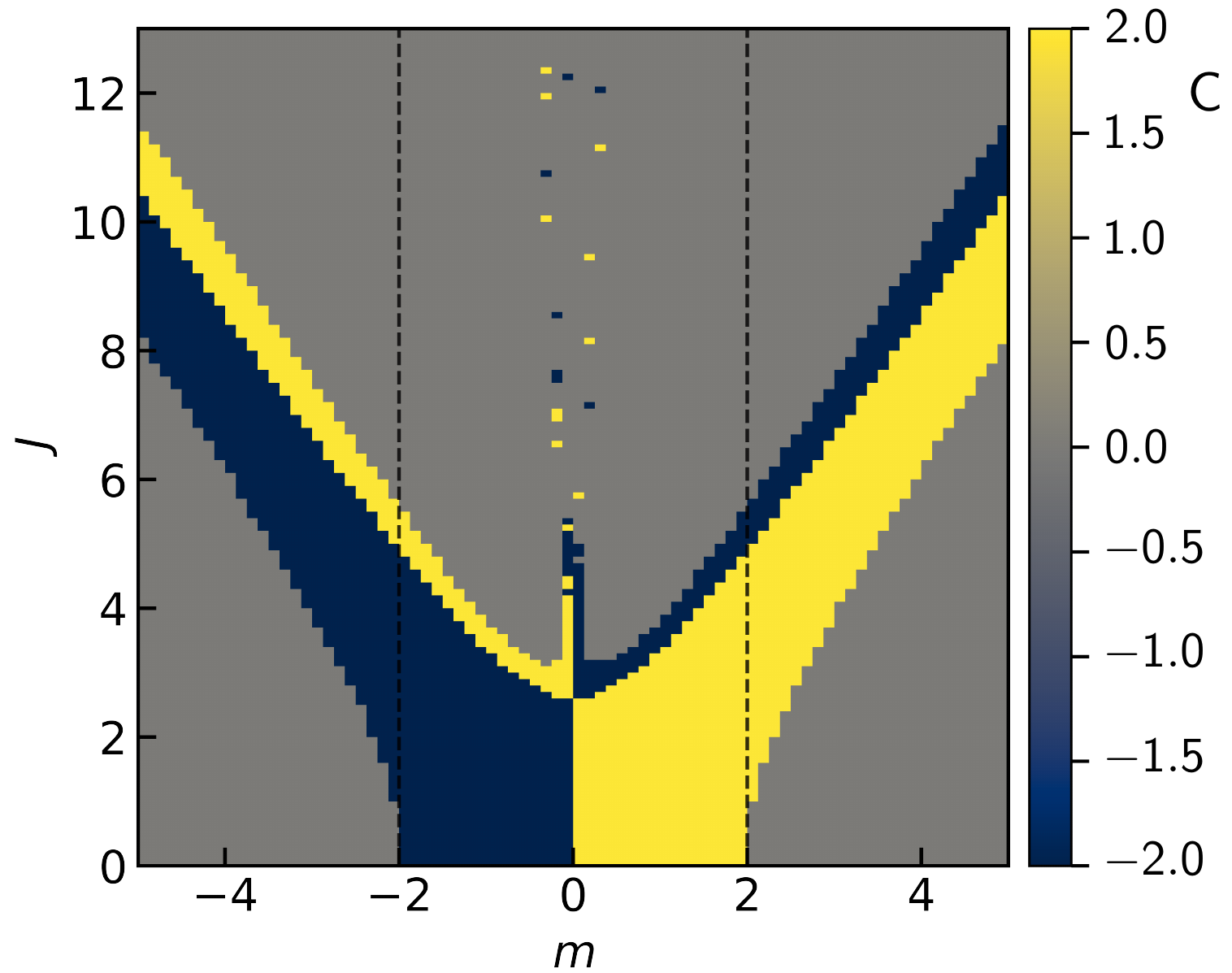}
\caption{
Topological phase diagram (as in Fig.\ \ref{fig:pd1}) but now obtained from the topological-Hamiltonian approach with disorder self-energy averaged of the BZ.
See the color code for the Chern number.
Results obtained for a system with $L_{x} \times L_{y} =8 \times 8$ sites with periodic boundary conditions. 
The disorder-averaged Green's function is obtained for $N_{\rm conf.} = 50$ disorder configurations. 
Chemical potential $\mu=0$.
$\eta=0.01$ in all calculations using the TH approach. 
}
\label{fig:pd3}
\end{figure}

Indeed, it turns out, that this is the dominant effect, which explains the structure of the phase diagram shown in Fig.\ \ref{fig:pd3}.
First, let us emphasize that the phase diagram, as obtained by the TH approach is qualitatively and even quantitatively the same as the one obtained by the TBC approach (Fig.\ \ref{fig:pd1}), given the uncertainties as specified by the standard deviation in Fig.\ \ref{fig:pd2} (upper panel). 
However, there is one exception, namely the scattered points around $m=0$ and for strong $J$ in Fig.\ \ref{fig:pd3}. 
These points indicate two features.
First, they indicate numerical artifacts, which primarily result from the coarse $m$-$J$ grid discretization used for the computations. 
Second, they point to potential signatures of physics beyond the TBC approach.
This requires further analysis, which is detailed below.

To start with, it is very instructive to consider the disorder self-energy in the atomic limit, i.e., for vanishing hopping $t=0$, as given by the local Hamiltonian \refeq{hloc}. 
We consider a certain configuration $S \in \mathbb{S}$ of classical spins, parameterized by spherical coordinates in the form $\ff S_{i\alpha} = (\sin \vartheta_{i\alpha} \cos \varphi_{i\alpha} , \sin \vartheta_{i\alpha} \sin \varphi_{i\alpha},\cos \vartheta_{i\alpha} )$. 
For a given disorder configuration $S$, the single-particle Green's function $G^{(S)}_{i\alpha\sigma, i'\alpha'\sigma'}(\omega)$ is readily obtained. 
It is diagonal in the site and orbital indices $G^{(S)}_{i\alpha\sigma, i'\alpha'\sigma'}(\omega) = \delta_{ii'} \delta_{\alpha\alpha'} G^{(\ff S_{i\alpha})}_{i \alpha, \sigma\sigma'}(\omega)$. 
For $\alpha = {\rm A}$, the $2 \times 2$ block matrix in the spin indices reads: 
\be
G^{(\ff S_{i\rm A})}_{iA}(\omega)
=
\begin{pmatrix}
\omega - m + \frac{J}{2} \cos \vartheta_{i\rm A} & \frac{J}{2} \sin \vartheta_{i\rm A} e^{-i \varphi_{i\rm A}} 
\\
\frac{J}{2} \sin \vartheta_{i\rm A} e^{i \varphi_{i\rm A}}  & \omega - m - \frac{J}{2} \cos \vartheta_{i\rm A}
\end{pmatrix}
^{-1}
\ee
while $G^{(\ff S_{i\rm B})}_{iB}(\omega)$ is obtained by the substitution $m \to -m$.
To perform the disorder average, it is sufficient to average the local Green's function at site $i$ over the unit sphere:  
$G_{iA}(\omega) = (4\pi)^{-1} \iint d \varphi_{i\rm A} d \vartheta_{i\rm A} \sin \vartheta_{i\rm A} G^{(\ff S_{i\rm A})}_{iA}(\omega)$. 
The atomic-limit disorder self-energy is obtained via Dyson's equation, $\ff \Sigma(\omega) = \ff G_{J=0}(\omega) ^{-1} - \ff G(\omega)^{-1}$, and turns out as diagonal with diagonal elements 
\be
\Sigma_{i\alpha\sigma}(\omega)
=
\frac{J^{2}}{4} \frac{1}{\omega \mp m}
\: ,
\labeq{alim}
\ee
and with the $-$ sign for $\alpha=\rm {A}$ and $+$ for $\alpha=\rm {B}$. 
This implies that the parameter $\Sigma$ in Eqs.\ (\ref{eq:sform}) and (\ref{eq:renor}) is given by:
\be
\Sigma = - \frac{J^{2}}{4} \frac{1}{m} \equiv \Sigma(m,J)
\: .
\labeq{sig}
\ee
We also note that \refeq{alim} is consistent with the sum rule 
\be
-\frac{1}{\pi} \int_{-\infty}^{\infty} d\omega \,  \mbox{Im} \,
\Sigma_{i\alpha\sigma, i'\alpha'\sigma'}(\omega+i \eta)
=
\delta_{ii'} \delta_{\alpha\alpha'} \delta_{\sigma\sigma'} \frac{J^{2}}{4}
\: ,
\labeq{sumrule}
\ee
which is {\em generally} valid (i.e., including $t \ne 0$) and which can be obtained straightforwardly by considering the high-frequency expansion of the disorder self-energy of the full model. 

Within the local approximation for the self-energy, 
the Chern number $C(m,J)$ for a fixed mass parameter $m$ and disorder strength $J$ is given in terms of the Chern number $C(m)$ of the clean system as
\be
C(m,J)
=
C(m^{\ast}(m,J))
\: , 
\labeq{renorm}
\ee
i.e., it does not depend on $J$ explicitly but only via the renormalized mass 
\be 
m^{\ast}(m,J) = m + \Sigma(m,J)
\labeq{mast}
\: . 
\ee
This implies that $C(m,J)$ can change its value discontinuously, if and only if the Chern number of the clean system, but with renormalized mass $m^{\ast}(m,J)$, changes discontinuously. 
For the clean system, the critical mass parameters are $m_{{\rm c},1}=-2$, $m_{{\rm c},2}=0$, and $m_{{\rm c},3}=+2$. 
For any $J>0$, the critical mass parameters $m = m_{{\rm c},r}(J)$ are thus obtained by solving 
the conditional equations $m^{\ast}(m,J) = m_{\rm c,r}$ for $m$, where $r=1,2,3$, i.e., by solving
\be
m + \Sigma(m,J) \stackrel{!}{=} m_{{\rm c},r}  
\labeq{cond}
\ee 
for $m$ for each $r=1,2,3$.

When additionally employing the atomic-limit approximation for the self-energy and using \refeq{sig}, the critical mass parameters are analytically  found as the solutions of \refeq{cond} with $\Sigma(m,J) = - J^{2} /4 m$, i.e., as the solutions of the quadratic equations 
$m - \frac{J^{2}}{4} \frac{1}{m} = m_{{\rm c},r}$.
This yields
\be
m_{{\rm c},r}(J) 
=
\frac{m_{{\rm c}, r}}{2} \pm \frac{J}{2} \sqrt{ 1+ \frac{ m_{{\rm c}, r}^{2} }{ J^{2} } } \quad  \mbox{for} \; r=1,2,3
\: .
\ee

If at all, the simple atomic-limit approximation applies to the strong-$J$ regime. 
Neglecting terms of order $m_{{\rm c}, r}^{2} / J^{2}$, we find critical mass parameters to be 
$m_{\rm c}(J) = \pm (J/2 - 1)$, $m_{\rm c}(J) =\pm J/2$, and $m_{\rm c}(J) = \pm (J/2 + 1)$.
These critical values nearly quantitatively explain the strong-$J$ regime of the phase diagram (see Fig.\ \ref{fig:pd3}).
Furthermore, we recover the correct sequence of Chern numbers as a function of increasing $m$: 
$C=$ $0$ / $-2$ / $+2$ / $0$ for $m<0$, and once more $C=$ $0$ / $-2$ / $+2$ / $0$ for $m>0$. 
This doubling of topological phases is controlled by the poles of the self-energy at $\omega = \pm m$ [see \refeq{sig}]. 
The strength of its residues, $J^{2}/4$, explains the asymptotically linear-in-$J$ shift of the nontrivial phases to ever-larger $|m|$. 

For weak $J$, the atomic-limit approximation breaks down even qualitatively. 
Although the pole structure of $\Sigma_{i\alpha\sigma}(\omega)$ [see \refeq{sig}] remains the same, there is only a {\em single} sequence of phases found numerically.
In the weak-disorder regime, the widening of the $C=\pm 2$ phases with $J$ indicates that the mass is still renormalized to smaller values, i.e., $\Sigma(m,J)<0$, see \refeq{mast} and \refeq{cond}.

There is a tetra-critical point at $m=0$ and $J\approx 2.5$ in the phase diagram shown in Fig.\ \ref{fig:pd3}, where four topological phases meet. 
This point separates the weak-disorder from the strong-disorder regime. 
Indeed, the two regimes are qualitatively different.
For $J \lesssim 2.5$, the $C=-2$ and $C=+2$ phases are separated at $m=0$ by a pole of the disorder-averaged Green's function at zero frequency. 
This pole evolves continuously from the zero-frequency pole of the averaged Green's function in the clean limit.
There it corresponds to the gap closure at the $X=(\pi,0)$ point in the Brillouin zone, which causes the semimetal phase.
For $J \gtrsim 2.5$, on the other hand, the $C=+2$ and $C=-2$ phases are separated at $m=0$ by a pole of the disorder self-energy 
at zero frequency, if the atomic-limit approximation is qualitatively correct.
This pole evolves continuously from the pole of the self-energy at $\omega=0$ for $m=0$ and $J\to \infty$, where the atomic approximation for the self-energy is reasonable.
Hence, for $m=0$ and with increasing $J$, the pole of the disorder-averaged Green's function at $\omega=0$ ceases to exist at the tetra-critical point and is replaced by a pole of the self-energy at $\omega=0$.

Note that a zero-frequency pole of the self-energy induces a zero of the Green's function at zero frequency.
This means that $\det \ff G(\omega) = 0$ for $\omega=0$ and implies a singularity of $\ff G(\omega)^{-1}$ in the Ishikawa-Matsuyama invariant given by \refeq{im}. 
Therefore, the Chern number can change as a function of $m$ at $m=0$. 
In the range $2.5 \lesssim J \lesssim 3.3$, the Chern number does change at $m=0$ in fact. 
Phase boundaries that emerge due to a zero of the Green's function are not possible in the clean ($J=0$) limit. 
Therefore, if the atomic-limit approximation qualitatively correctly captures the physics, then 
the phase transition from $C=+2$ to $C=-2$ at $m=0$ in this $J$ range and, therefore, the $C=+2$ and $C=-2$ phases represent novel phases, which have no counterpart in the clean limit, since the clean-limit Green's function cannot have zeros.

\begin{figure}[t] 
\centering
\includegraphics[width = 0.85\linewidth]{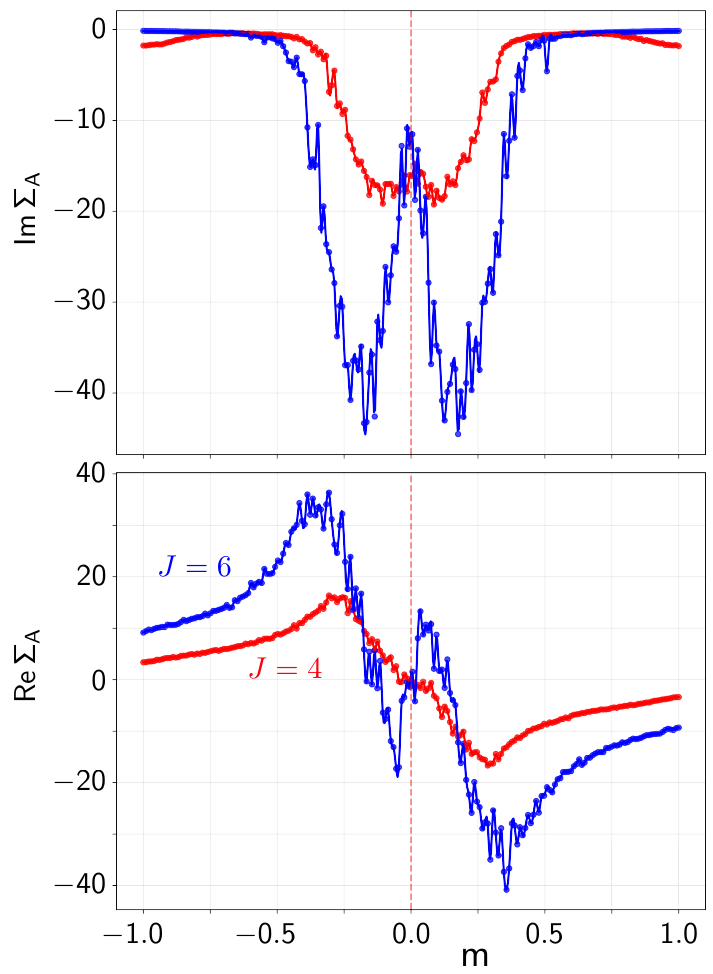}
\caption{
Imaginary (upper panel) and real part (lower panel) of the retarded ($\ff k$-averaged) disorder self-energy at $\omega=0$ for $\alpha = {\rm A}$ as function of $m$ in the range $-1 \le m \le 1$. 
{\it Red}: $J=4$, {\it blue}: $J=6$.
Results as obtained for a system with $10 \times 10$ sites, and with $N_{\rm conf.} = 200$ disorder configurations.
Lorentzian broadening $\eta=0.01$.
}
\label{fig:sig0}
\end{figure}

The decisive question is whether the above-described scenario is merely an artifact of the rather crude atomic-limit approximation.
In a first step, we have therefore computed the full but still local disorder self-energy at $\omega = 0$ numerically in a very narrow range around $m=0$.
Fig.\ \ref{fig:sig0} shows the imaginary and the real parts of the A component of 
$\Sigma_{\rm A} = \Sigma_{i {\rm A}\sigma}(i\eta)$, i.e., at $\omega=0 + i \eta$, for a small broadening parameter $\eta=0.01$ and as a function of $m$.
For the B orbital, we find $\Sigma_{\rm B} = - \Sigma_{\rm A}$ (not shown). 
The $m$ dependence of $\Sigma_{\rm A}$ is found to be qualitatively different for $J \gtrsim 5$ and $J \lesssim 5$. 

For strong disorder ($J \gtrsim 5$, see data for $J=6$ in the figure) there are two negative peaks of $\mbox{Im} \, \Sigma_{A}$ as a function of $m$, symmetrically located around $m=0$.
This two-peak structure differs from the atomic-limit approximation, \refeq{alim}, which predicts $\mbox{Im}\, \Sigma_{\rm A} = - (\pi J^{2}/4) \delta(m)$, i.e., a single delta peak at $m=0$.
However, it is plausible that a weak but a non-vanishing hybridization between the A and B orbitals emerges at the order $t^{2}/J$ in a perturbative expansion around the atomic limit. 
This hybridization is expected to produce a splitting $\pm \Delta m$ of the single pole at $m=0$ on a scale $\Delta m \sim t^{2}/J = 0.167$ at $J=6$. 
In fact, this estimate compares well with the splitting between the two peaks of $\mbox{Im} \, \Sigma_{A}$ seen in the figure.

For intermediate disorder strengths ($2.5 \lesssim J \lesssim 5$), see the data for $J=4$ in the figure, there is a single broader peak of $\mbox{Im} \, \Sigma_{A}$ centered around $m=0$.
In the narrow $m$ range between $m \approx \pm 0.5$, the real part of the local self-energy $\Sigma_{\rm A}$ at $\omega=0$ first strongly increases with increasing $m$, reaching a maximum $\mbox{Re} \, \Sigma_{\rm A} \approx 18$ at $m \approx -0.3$, and then rapidly decreases to zero at $m=0$. 
Following the symmetry relation $\mbox{Re}  \,\Sigma_{A}(m) = - \mbox{Re} \, \Sigma_{A}(-m)$, this behavior continues for $m>0$. 
Within the local approximation (see \refeq{sform}), we have $\mbox{Re} \, \Sigma_{A}(m) = \Sigma$, and we can therefore argue as above and use the resulting mass renormalization $m^{\ast} = m + \Sigma$ to find the Chern number from \refeq{renorm}.
This yields another sequence of phases with $C=$ $0$ / $+2$ / $-2$ / $0$ emerging close to and centered around $m=0$, where $m=0$ marks the boundary between phases with $C = + 2$ and $C=-2$.
Compared with the clean-limit phase diagram, note that the sequence is inverted due to the strictly {\em decreasing} $\mbox{Re}\, \Sigma_{\rm A}$ in the range $-0.3 \lesssim m \lesssim 0.3$.
This inverted sequence is exactly what is seen in the phase diagram (Fig.\ \ref{fig:pd3}) at $J=4$, but only a few pixels represent the very narrow $m$ range.

For $J \gtrsim 5$ (see $J=6$ in Fig.\ \ref{fig:sig0}) the situation is more complicated due to the additional structure in $\mbox{Re} \, \Sigma_{\rm A}(m)$ in the even narrower range $-0.05 \lesssim m \lesssim 0.05$, resulting in three additional sequences of $C$ values rather than one in the range $-0.3 \lesssim m \lesssim 0.3$. 
As $m$ increases, we find the sequence $C=$ $0$ / $+2$ / $-2$ / $0$ for $m<0$.
This sequence is roughly centered around the first zero-crossing of $\mbox{Re}\, \Sigma_{\rm A}$ and inverted as compared to the clean-limit phase diagram due to the strictly {\em decreasing} $\mbox{Re}\, \Sigma_{\rm A}$.
Next is the sequence $C=$ $0$ / $-2$ / $+2$ / $0$, centered around the second zero-crossing at $m=0$ and in the same order as for $J=0$ but in the tiny $m$ range, where $\mbox{Re}\, \Sigma_{\rm A}$ strictly increases with $m$.
Finally, there is another (inverted) sequence $C=$ $0$ / $+2$ / $-2$ / $0$ for $m>0$ around the third zero-crossing.
Note that these sequences cannot be resolved with the discretization of the $m$ and $J$ grids used for the overview phase diagram in Fig.\ \ref{fig:pd3}.

\section{Non-locality of the disorder self-energy}
\label{sec:nonloc}

Finally, we discuss the $\ff k$ dependence of the disorder self-energy, or its non-locality, which has been neglected so far, see \refeq{kindep}. 
At $\omega=0$ the $\ff k$ dependence leads to a $\ff k$-dependent renormalization of the mass parameter and thus to a $\ff k$-dependent deformation of the topological Hamiltonian \refeq{toph}, i.e., a more complex deformation beyond a simple single-parameter renormalization. 

The strength of the $\ff k$ dependence of the self-energy depends sensitively on the orbital polarization, which is defined in terms of the average site occupations $n_{i\alpha} = \sum_{\sigma=\uparrow,\downarrow} \langle c^{\dagger}_{i\alpha\sigma} c_{i\alpha\sigma} \rangle$ as
\be
p = \frac12 (n_{i \rm B} - n_{i\rm A})
\: . 
\ee
A strong polarization $p \to 1$ implies that the A (B) orbitals are nearly unoccupied (fully occupied) and vice versa for $p\to -1$.
Therefore, the ground-state expectation value of the local electron spin $\langle \ff s_{i\alpha} \rangle$ on an $\alpha=\rm A$ (B) orbital becomes small such that the coupling term \refeq{dis} is less active. 
Even for strong $J$, this leads to an effectively weak disorder and thus to a nearly local self-energy.
 
\begin{figure}[t] 
\centering
\includegraphics[width = 0.8\linewidth]{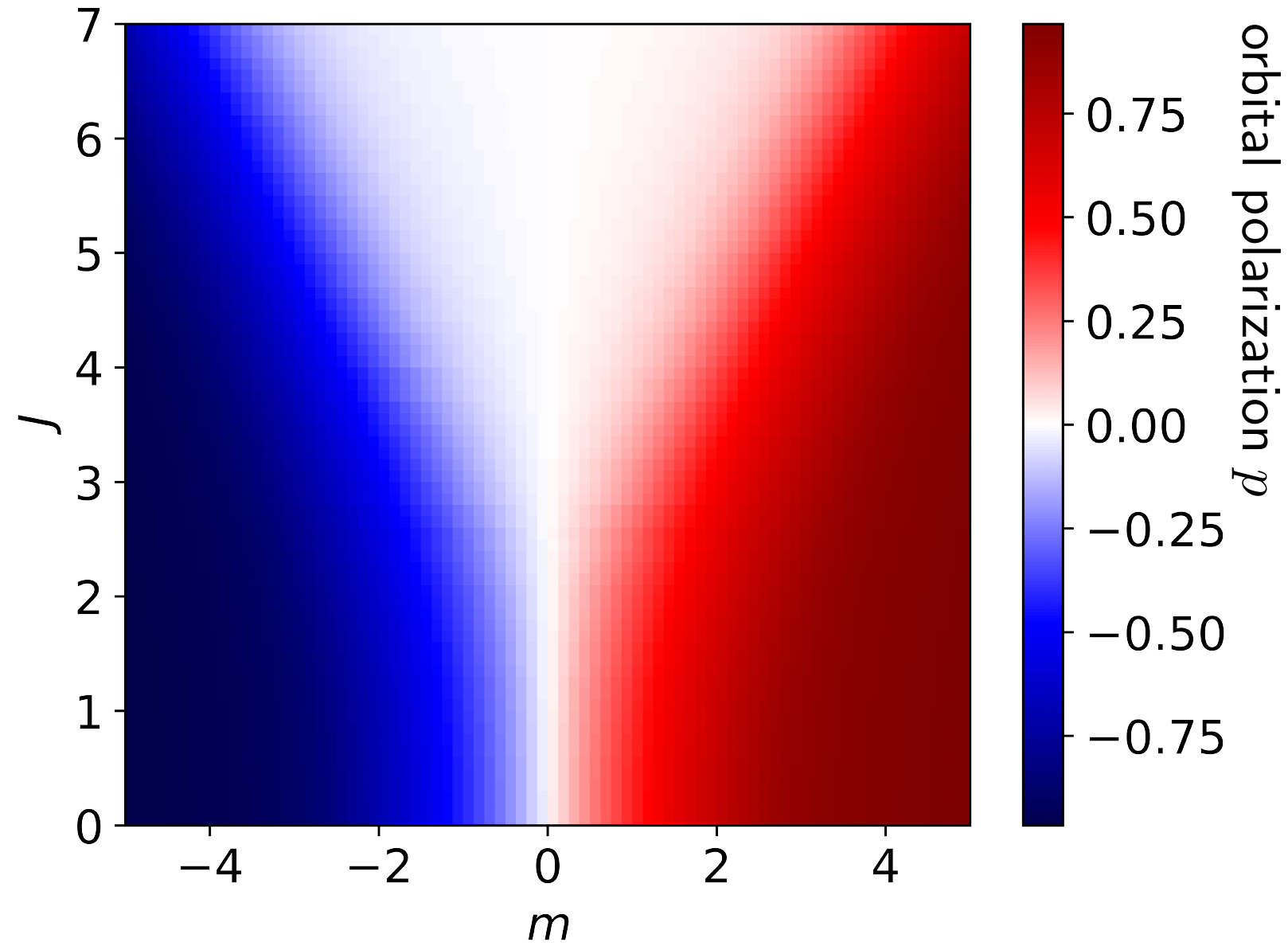}
\caption{
Average orbital polarization $p=\frac12 (n_{i \rm B} - n_{i\rm A})$ in the $m$-$J$ plane. 
Results obtained for $L=16 \times 16$ and $N_{\rm conf.}=1$ and the full, $\ff k$-dependent self-energy.
}
\label{fig:pol}
\end{figure}

The polarization $p$ is controlled by the mass parameter $m$. 
Figure \ref{fig:pol} demonstrates that full polarization, $p \to \pm 1$, is obtained for $m \to \pm \infty$ for any fixed $J$ and that with increasing $m$ the orbital polarization smoothly increases. 
For strong $m$ and $J$ with $J=\pm 2m$, we find $p \approx \pm 0.5$. 
Importantly, $p=0$ for $m=0$, irrespective of $J$, and furthermore the polarization almost vanishes in an ever larger $m$ interval around $m=0$ with increasing $J$.
Hence, in this regime, we expect a possibly significant $\ff k$ dependence and corresponding changes of the phase boundaries.

\begin{figure}[t] 
\centering
\includegraphics[width = 0.99\linewidth]{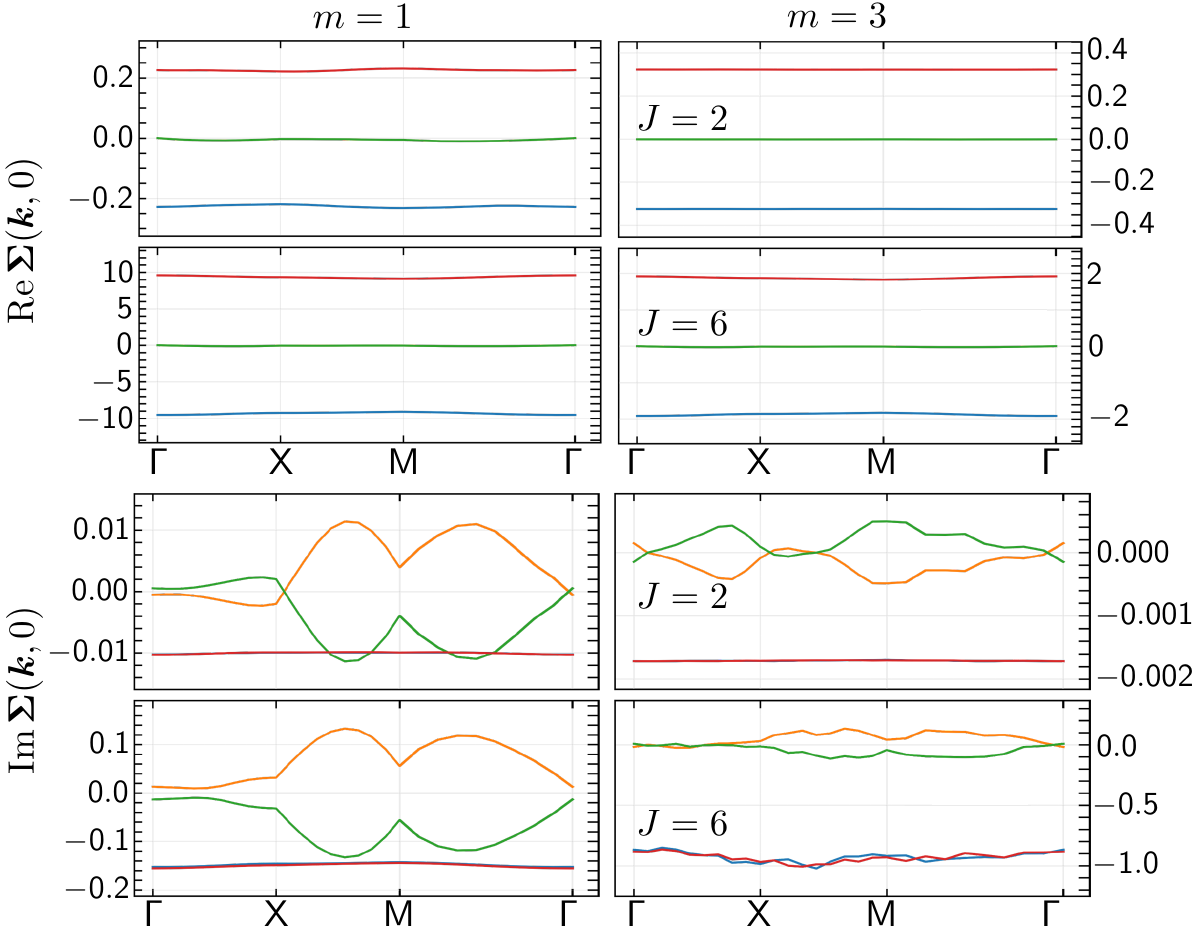}
\caption{
Real part (upper panels for $J=2$ and $J=6$) and imaginary part (lower panels for $J=2$ and $J=6$) of the disorder self-energy $\ff \Sigma(\ff k,\omega)$ at $\omega=0$ along high-symmetry directions in the two-dimensional Brillouin zone.
{\em Left panels:} $m=1$.
{\em Right panels:} $m=3$.
{\em Blue lines}: AA component of the self-energy. 
{\em Red}: BB. 
{\em Green/orange}: off-diagonal AB/BA components. 
Results as obtained for $L=20 \times 20$, $N_{\rm conf.}=300$, and $\eta=0.01$.
}
\label{fig:sigmare}
\end{figure}

Figure \ref{fig:sigmare} shows the $\ff k$ dependence of the real part (upper panels) and the imaginary part (lower panels) of the retarded self-energy at $\omega=0$. 
We first concentrate on the real part.
Here the orbital-diagonal (AA and BB) components are strongly dominating for all $m$, $J$ combinations.
The effect of orbital polarization on the $\ff k$ dependence becomes obvious by comparing the results for $m=1$ with those for $m=3$ at a weak disorder strength $J=2$. 
For $m=1$ there is a finite $\ff k$ dependence of $\mbox{Re}\, \ff \Sigma(\ff k, 0)$ for both, the intra-orbital (AA, BB) as well as of the inter-orbital (AB / BA) components. 
Contrary, for $m=3$ the $\ff k$ dependence of $\mbox{Re}\, \ff \Sigma(\ff k, 0)$ is negligibly small. 
This correlates with the much stronger polarization at $m=3$, see Fig.\ \ref{fig:pol}.
For strong disorder, $J=6$, the polarization is not strong enough in both cases, $m=1$ and $m=3$, to suppress the $\ff k$ dependence of $\mbox{Re}\, \ff \Sigma(\ff k, 0)$. 

The strength of the $\ff k$ dependence can be quantified by the difference $\Delta \ff \Sigma$ between the maximum and the minimum values of the components of $\mbox{Re}\, \ff \Sigma(\ff k, 0)$ along the high-symmetry directions in the Brillouin zone.
We find that the {\em relative} strength of the $\ff k$ dependence, i.e., $\Delta \ff \Sigma$ compared to the $\ff k$-averaged values $\mbox{Re}\, \ff \Sigma(\ff k, 0)$, component by component, is almost the same in all cases, except for $m=3$, $J=2$, where the $\ff k$ dependence is nearly vanishing due to the strong polarization.

As demonstrated with the lower panels of Fig.\ \ref{fig:sigmare}, the imaginary parts of the components of the self-energy $\mbox{Im}\, \ff \Sigma(\ff k, 0)$ follow the same trends as the reals parts. 
We also note that the inter-orbital components sum up to zero. 

\begin{figure}[t] 
\centering
\includegraphics[width = 0.85\linewidth]{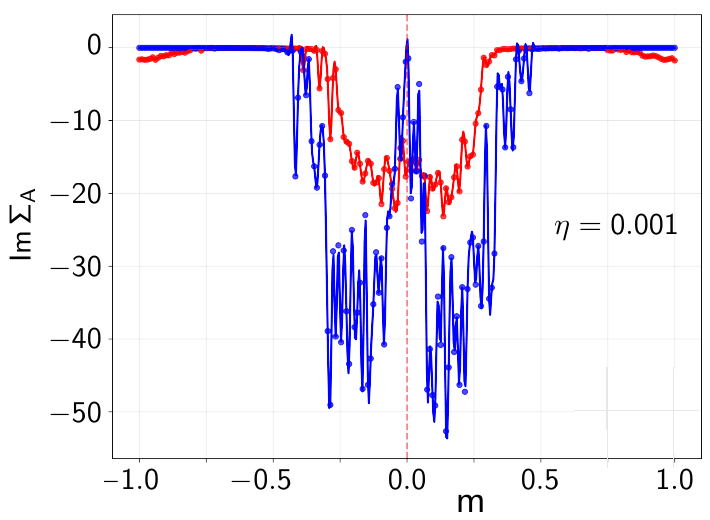}
\caption{
Imaginary part of the self-energy for $J=4$ (red) and $J=6$ (blue) as in Fig.\ \ref{fig:sig0} but f\"ur $\eta=0.001$.
}
\label{fig:sig0eta}
\end{figure}

For the model with two orbitals A and B per site, the $\ff k$-dependent disorder self-energy is a $2 \times 2$ matrix $\ff \Sigma(\ff k, \omega)$ and, thus, there are two bands of self-energy poles $\omega_{j}(\ff k)$ with $j=1,2$.
Since a pole of the self-energy implies a zero of the disorder-averaged Green's function at the same frequency $\omega_{j}(\ff k)$, we refer to them as zero bands, a concept that is well known from correlated electron systems \cite{Gur11,WCA+23}.
The width of the single peak for $J=4$ and the widths of the two peaks for $J=6$ of $\mbox{Im} \, \Sigma_{A}$, see Fig.\ \ref{fig:sig0}, is only to a marginal extent induced by the damping constant $\eta$. 
This is demonstrated by repeating the computation of $\mbox{Im} \, \Sigma_{A}$ with $\eta = 0.001$, i.e., a damping constant that is by a factor 10 smaller than the one used to get the results shown in Fig.\ \ref{fig:sig0}.
As is shown in Fig.\ \ref{fig:sig0eta}, the peak widths are basically unaffected by the smaller $\eta$, and only the scattering of the data increases.

From the discussion of the local approximation for the self-energy (see Sec.\ \ref{sec:resth}), where the zero-bands are flat with $\omega_{\rm A}(\ff k) = m$ and $\omega_{\rm B}(\ff k) = -m$, we know that the Chern number changes from $C=+2$ to $C=-2$ with increasing $m$ at $m=0$, where both flat zero-bands cross $\omega=0$. 
In the dispersive case, a crossing of the zero-bands at $\omega=0$ must therefore, by continuity, lead to the same change of the Chern number, at least in the simpler case $J=4$, where the real part is strictly decreasing function of $m$ close to $m=0$.
This transition is still expected to take place at $m=0$ because of the particle-hole symmetry of the system.
We strongly emphasize that a crossing of {\em poles} of the Green's function cannot be responsible for the topological transition discussed here, since in the relevant small $m$ interval around $m=0$ at $J=4$ there is a well-developed spectral gap. 
This can be seen in Fig.\ \ref{fig:spec}: 
There are no states with eigenenergies and thus no poles of the Green's function at $\omega=0$ in an extended interval $-1 < m < 1$ around $m=0$.

For the accessible system sizes, the $\ff k$-dependent zero-band energies as function of $m$ cannot be reliably obtained numerically. 
Hence, we cannot exclude {\em a priori} that the transition from $C=+2$ to $C=-2$ with increasing $m$ takes place via a small but finite range $\Delta m$ around $m=0$, where the Chern number is ill-defined due to overlapping zero-bands.
Let us emphasize that in this case the transition would still be driven by the zero-bands rather than by the poles of the Green's function and is thus {\em qualitatively} different from transitions in the clean limit.

To check our expectation and, more generally, to study the impact of the $\ff k$ dependence of the self-energy on the phase boundaries in the topological phase diagram, we compute the Chern number from the topological Hamiltonian with $\ff k$-dependent renormalization.
We start from \refeq{toph} with the full nonlocal disorder self-energy $\mbox{Re} \, \Sigma_{i\alpha\sigma, i' \alpha',\sigma'}(\omega=i\eta)$, with a small imaginary damping constant $\eta = 0.01$.
Fourier transformation yields the $\ff k$-dependent self-energy and therewith the topological Hamiltonian 
$H_{\rm top} = \sum_{\ff k, \alpha\alpha', \sigma} H_{\rm top}(\ff k)_{\alpha\alpha'} c^{\dagger}_{\ff k\alpha\sigma} c_{\ff k\alpha'\sigma}$, where the $2 \times 2$ Bloch Hamiltonian is brought into the form 
$\ff H_{\rm top}(\ff k) = \ff \epsilon(\ff k) + \ff \Sigma(\ff k, i \eta) \equiv \sum_{\ff k} \ff d (\ff k) \cdot \ff \tau$.
The Chern number can then be obtained via an integral \cite{QWZ06}
\be
C = \frac{-1}{4\pi} \iint_{\rm BZ} dk_{x}dk_{y} \, \ff d(\ff k) \cdot \partial_{k_{x}} \ff d(\ff k) \times \partial_{k_{y}} \ff d(\ff k) \: 
\labeq{ch}
\ee
over the Brillouin zone. 

\begin{figure}[t] 
\centering
\includegraphics[width = 0.8\linewidth]{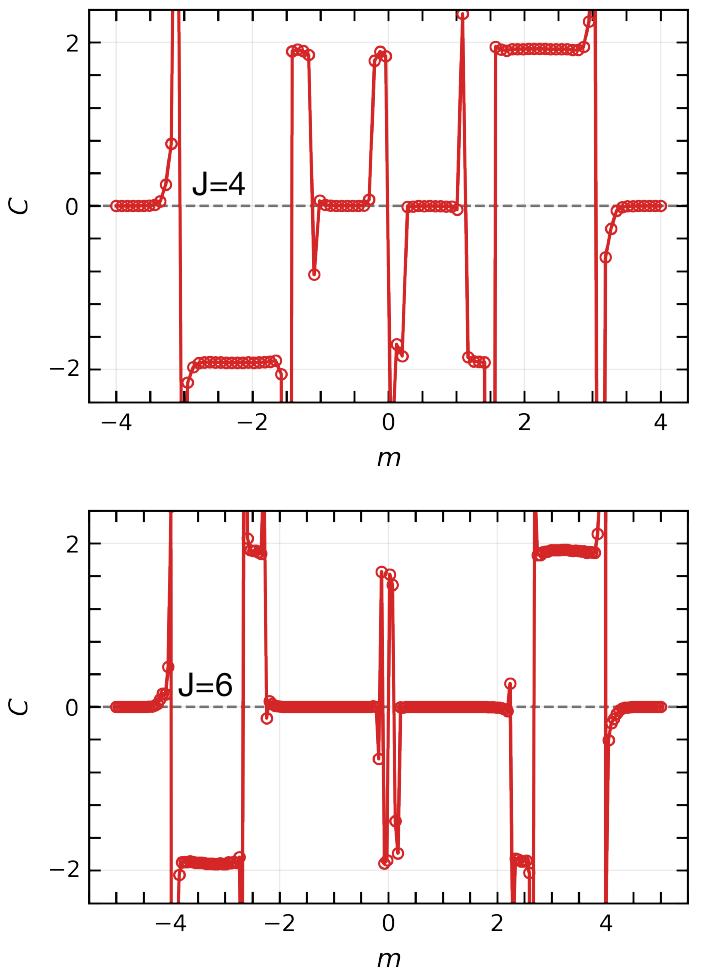}
\caption{
Chern number as a function of $m$ for $J=4$ (upper panel) and $J=6$ (lower panel), as obtained from \refeq{ch} for a lattice with $L = 18 \times 18$ sites, $N_{\rm conf.} = 200$ disorder configurations, and $\eta=0.01$.
}
\label{fig:ckdep}
\end{figure}

The resulting Chern number as a function of $m$ for $J=4$ and $J=6$ is shown in Fig.\ \ref{fig:ckdep}.
Deviations of $C$ from integer values $C=-2,0,2$ indicate computational issues due to the limited system size $L=18\times 18$ and due to the necessity to compute the partial derivatives in $k_{x}$ and $k_{y}$ directions in \refeq{ch} numerically.
A first important insight concerns the transitions at larger mass parameters $|m|$ seen in Fig.\ \ref{fig:ckdep}:
For $J=4$ (upper panel), the critical mass parameters, with discontinuous jumps of $C$ in the $m>0$ range, are the same as those in Fig.\ \ref{fig:pd3}, where 
$m_{\rm c} \approx 3.0$ (boundary between phases I and II),
$m_{\rm c} \approx 1.5$ (II and III), 
$m_{\rm c} \approx 1.2$ (III and IV) within an uncertainty of $\Delta m \approx 0.1$. 
At $J=6$, with the same uncertainty, agreement is found with 
$m_{\rm c} \approx 3.9$ (boundary between phases I and II),
$m_{\rm c} \approx 2.7$ (II and III), 
$m_{\rm c} \approx 2.2$ (III and IV). 
We conclude that consideration of the $\ff k$ dependence of the self-energy might only slightly shift the phase boundaries in this regime.

\begin{figure}[t] 
\centering
\includegraphics[width = 0.8\linewidth]{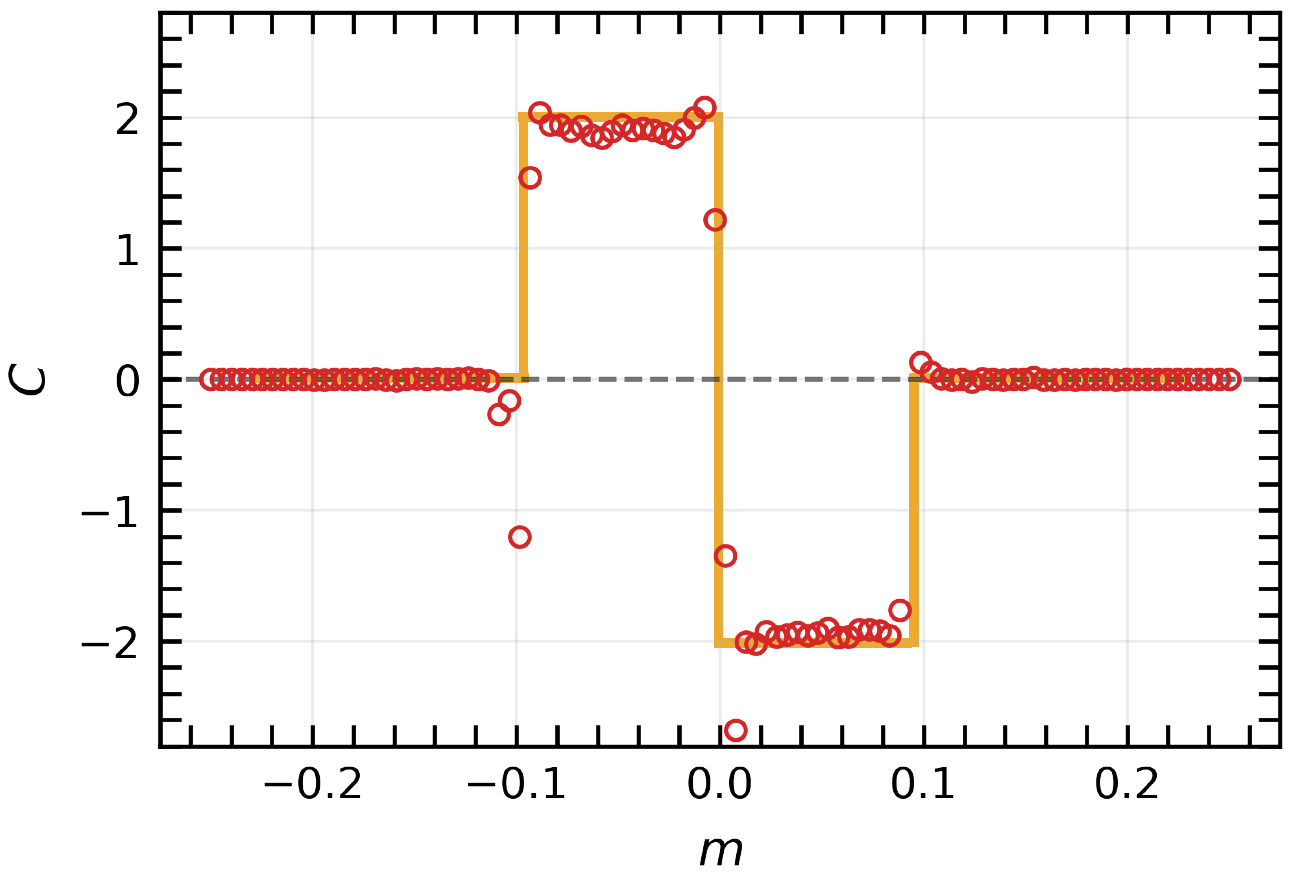}
\caption{
Chern number as a function of $m$ for $J=4$, obtained from the full $\ff k$-dependent self-energy as in Fig.\ \ref{fig:ckdep}, but in the range $-0.25 < m <0.25$ and with $N_{\rm conf.} = 400$, $\eta=0.1$.
Yellow lines $C=0,\pm 2$ are guides to the eye.
}
\label{fig:ckdep4}
\end{figure}

Close to $m=0$, the resolution in Fig.\ \ref{fig:pd3} is insufficient for a meaningful comparison. 
Furthermore, for $J=6$ (see Fig.\ \ref{fig:sig0}) and stronger $J$, there is strongly non-monotonic behavior of the real part of the local disorder self-energy in a very small range around $m=0$. 
As discussed above, within the local approximation, this leads to a multitude of topological transitions in the same small $m$ interval. 
Resolving this structure at $J=6$ in a full calculation based on the $\ff k$-dependent self-energy, is highly demanding and would require a more elaborate computational approach. 
Contrary, at $J=4$, the real part of the local self-energy in the vicinity of $m=0$ is a strictly decreasing function of $m$, see Fig.\ \ref{fig:sig0}, and the local approximation predicts a single sequence of Chern numbers $C=$ $0$ / $+2$ / $-2$ / $0$ only.
The full calculation, see upper panel of Fig.\ \ref{fig:pd3}, is in fact consistent with this result. 
Calculations with a better resolution are shown in Fig.\ \ref{fig:ckdep4}.
While finite-size effects are still non-negligible, we can safely confirm the qualitative implications of the local approximation for the sequence of Chern numbers with critical mass parameters given by 
$m_{\rm c} \approx -0.095$ ($C=0 \to +2$), 
$m_{\rm c} \approx 0$ ($C=+2 \to -2$), 
and 
$m_{\rm c} \approx +0.095$ ($C=-2 \to 0$), 
with an uncertainty of $\Delta m \approx 0.02$. 

We conclude that the disorder self-energy exhibits a significant $\ff k$ dependence, except for regimes in the phase diagram, where the orbital polarization is strong.
In a tiny interval $-0.3 \lesssim m \lesssim 0.3$ and for strong $J$, the $\ff k$-dependent self-energy has poles close to $\omega=0$ that give rise to two dispersive Green's-function zero-bands $\omega_{1,2}(\ff k)$. 
They are related to the two non-dispersive poles of the self-energy in the atomic limit but develop an energy splitting at $m=0$ due to the nonlocal hybridization between the A and B orbitals. 
As a function of $m$, these zero-bands are crossing the chemical potential and lead to topological transitions between phases characterized by different Chern numbers.
This implies that the transition between phase III for $m<0$ ($C=+2$) and phase III for $m>0$ ($C=-2$) is qualitatively different from phase transitions in the clean limit. 
Hence, phase III itself (for $m<0$ and $m>0$) is a topologically nontrivial phase, where the topology is carried by Green's-function zeros. 
Since this is impossible in the case of a clean-limit Green's function, 
phase III represents a novel phase that is {\em induced} by disorder and that might be called a topological Anderson insulator.

\section{S-space Chern number}
\label{sec:space}

The phases I and IV in the phase diagram shown in Fig.\ \ref{fig:pd1} are both ``topologically trivial'' in the sense that $C=0$. 
Recall that the Chern number $C$ is the usual invariant for two-dimensional non-interacting systems in the Altland-Zirnbauer class D \cite{AZ97,RSFL10}. 
It topologically characterizes the bundle of Bloch states over the first Brillouin zone, i.e., over the two-dimensional manifold of wave vectors $\ff k$.
In addition to this ``k-space topology,'' the presence of the manifold $\mathbb{S}$ of classical spin configurations $S \in \mathbb{S}$ allows an independent topological characterization. 
Analogous to the k-space Chern number $C$, we can compute the S-space Chern number $C^{\rm (S)}$, which has been introduced recently \cite{MFQ+24} for tight-binding systems containing a single or a few classical impurity spins. 
For systems with $R$ classical spins coupled to the electron system via, e.g., an exchange interaction of the form \refeq{dis}, the closed and simply connected parameter manifold $\mathbb{S} = \mathbb{S}^{2} \times \cdots \times \mathbb{S}^{2}$ is $2R$ dimensional. 
If the electron system has a gapped ground state $| \Psi_{0}\rangle$ on $\mathbb{S}$, the $R$-th S-space Chern number, defined via
\be
  C^{\rm (S)}_{R} = \frac{i^{R}}{(2\pi)^{R}} \frac{1}{R!} \oint_{\ca S} \mbox{tr} \, \Omega^{R} \: ,
\labeq{spinchern}
\ee
can only have integer values, $C^{\rm (S)}_{R} \in \mathbb{Z}$ \cite{Nak98,MP22}.
Here, $\Omega = dA$ is the S-space Berry-curvature two-form derived from the S-space Berry connection one-form $A$ of the ground-state bundle over $\mathbb{S}$.

In two spatial dimensions the ground state of a disordered system is generically given by a Slater determinant formed by {\em localized} one-particle eigenstates.
In this case, the $R$-th S-space Chern number is invariant under continuous deformations of the electronic Hamiltonian as long as there is no gap closure due to a delocalized state for all spin configurations $S \in \mathbb{S}$.
For the model considered here, we have $R=2L$. 
Upon parameterizing the spin configuration $S = (\ff S_{1A}, \ff S_{1B}, ..., \ff S_{LA}, \ff S_{LB})$ by polar and azimuthal angles $\ff \lambda = (\lambda_{0}, ... , \lambda_{2R-1}) \equiv (\vartheta_{0}, \varphi_{0}, ... , \vartheta_{R-1}, \varphi_{R-1})$, we obtain
\ba
C^{\rm (S)}_{R} &=& \frac{i^{R}}{(2\pi)^{R}} \frac{1}{R!} 
\sum_{\pi} 
\mbox{sign}\, {\pi}
\int d\lambda_{0} \cdots d\lambda_{2R-1}
\nonumber \\
&&
\frac{\partial \langle \Psi_{0} |}{\partial \lambda_{\pi(0)}}
\frac{\partial | \Psi_{0} \rangle}{\partial \lambda_{\pi(1)}}
\cdots
\frac{\partial \langle \Psi_{0} |}{\partial \lambda_{\pi(2R-2)}}
\frac{\partial | \Psi_{0} \rangle}{\partial \lambda_{\pi(2R-1)}}
\: ,
\labeq{spinchernang}
\ea
where the sum runs over all permutations $\pi$, see Ref.\ \cite{MFQ+24} for details.

There are two limit cases where the S-space Chern number can be determined easily. 
(i) In phase I for $J=0$ and $|m| >2$, we trivially have $C^{\rm (S)}_{2L} = 0$, since all spins are decoupled. 
Thus, phase I, at $J=0$ and also for $J < J_{\rm c1}(m)$, is S-space topologically trivial.
(ii) In phase IV, in the limit $J \to \infty$ for any fixed $m$, the atomic-limit Hamiltonian $H_{\rm local}$ given by \refeq{hloc} captures the S-space topology.
$H_{\rm local} = \sum_{i\alpha} H_{i\alpha}$ describes a system of decoupled magnetic-monopole models \cite{Dir31,Ber84,Sim83}
\be
H_{i\alpha}
= 
m 
z_{\alpha} n_{i\alpha}
+
J
\ff s_{i\alpha} \ff S_{i\alpha} 
\: ,
\labeq{mono}
\ee
which include an additional, but topologically irrelevant, orbital-dependent but spin-independent energy with a sign factor defined as $z_{A}=1$ and $z_{B}=-1$.
Since $H_{\rm local}$ is additive, the $R$-th S-space Chern number ($R=2L$) of this effective model factorizes into a product of the first S-space Chern numbers $C^{\rm (S)}_{1,i\alpha}$ of each individual magnetic monopole. 
For $J>0$, we have $C^{\rm (S)}_{1,i\alpha} = +1$ and thus
\be
C^{\rm (S)}_{2L} = \prod_{i=1}^{L} \prod_{\alpha=\rm A, B} C^{\rm (S)}_{1,i\alpha} = 1
\: .
\ee
We conclude that phases I and IV have the same k-space but different S-space Chern numbers and thus cannot be deformed into each other continuously.

\section{Concluding discussion}
\label{sec:con}

The QWZ model is a generic tight-binding model of a Chern insulator. 
We have analyzed the topological phase diagram of this model in the presence of an exchange coupling to disordered local magnetic moments, classical spins of unit length, by employing two complementary numerical approaches -- the twisted-boundary-condition (TBC) method and the topological-Hamiltonian (TH) approach -- as well as analytical considerations.
As is well known, disorder in two-dimensional systems leads to localization of all single-particle eigenstates, except at certain isolated critical energies which act as critical points between the localized states.
The TBC approach makes explicit that the appearance of {\em localized} states at the chemical potential does not affect the topology of the bulk system, such that topological invariants, in particular the Chern number in the present case, remain well defined.

The topologically nontrivial phases of the clean system turn out to be robust against weak disorder. 
For fixed small exchange-coupling strength $J$, both the TBC and the TH approaches predict, as a function of $m$, the familiar sequence of phases with Chern numbers $C=0$, $C=-2$, $C=+2$, and back to $C=0$, with a slightly enlarged mass-parameter window for the $C= \pm 2$ phases.

More remarkably, we also find nontrivial topological phases for strong magnetic disorder $J$. 
This is counterintuitive because for systems with {\em fixed} classical spin textures strongly coupled to itinerant electrons, nontrivial topology requires finite scalar spin chirality, $\chi_{ijk} = \ff S_{i} \times \ff S_{j} \cdot \ff S_{k} \ne 0$ \cite{YKM+99,OMN00,AM10,IN18,HDKZ20,TR21,DLPP22,Hay23,KKF23,ZLLZ25}. 
For spin {\em disorder}, however, the average chirality vanishes, implying no net Berry curvature from the spin background. 
At the same time, while the disorder averages to an effective restoration of time-reversal symmetry, this symmetry remains explicitly broken by the clean QWZ Hamiltonian. 
Hence, there is no symmetry-based obstruction to disorder-induced topology in this system.

A key result of our study is that nontrivial phases persist for {\em arbitrarily strong disorder} in the combined limit $J\to \infty$ and $m \to \pm \infty$.
Specifically, the sequence $C=$ $0$ / $-2$ / $+2$ / $0$ appears twice as a function of $m$: once near $m=-J/2$ and once near $m=+J/2$, i.e., for $\delta = m \pm J/2 = \ca O(t)$. 
Both approaches yield identical Chern numbers and critical mass parameters $m_{\rm c}(J)$, demonstrating the reliability of the phase boundaries.

The underlying mechanism becomes transparent in the atomic limit. 
The positions of the nontrivial phases in the $m$-$J$ plane reflect the competition between spin polarization (set by $J$) and orbital polarization (set by $m$). 
The effective low-energy Hamiltonian, \refeq{heff}, predicts the $m$-dependent boundaries between the $C=+2$ and $C=-2$ phases with quantitative accuracy and directly identifies the delocalized state at the chemical potential that drives each topological transition.

When the disorder self-energy is approximated as local, the TH approach effectively maps the phase boundaries in the strongly disordered regime onto those of the clean system.
Thus, the strong-disorder topology is still rooted in the band inversion of the clean QWZ model.
In this regime, simple analytic expressions for the boundaries emerge and agree very well with numerical results.

A central conceptual question that remains is whether disorder can generate topologically nontrivial phases that are {\em genuinely} absent in the clean system. 
This issue has been debated extensively \cite{LCJS09,GWA+09,JWSX09,YNIK11,Pro11a}, but so far answered negatively.
A convincing demonstration requires showing either (i) that the phase possesses a topological invariant not realizable in the clean model, or, if the invariants coincide, that (ii) there exists a strict argument proving that the phase is qualitatively different and cannot be realized in the clean limit.
In the present work, we provide explicit positive examples for both strategies.

For strategy (i), we exploit the fact that the many-body ground state defines a vector bundle over the space of classical spin configurations $\mathbb{S} = \mathbb{S}^{2} \times \cdots \times \mathbb{S}^{2}$.
One may thus characterize topology not only via the conventional momentum-space Chern number but also via an ``S-space'' Chern number, in particular the $2L$-th spin Chern number $C^{\rm (S)}_{2L}$.
Although $\dim \mathbb{S} = 4L$ renders a numerical evaluation prohibitively difficult, analytic considerations in the clean and atomic limits yield $C^{\rm (S)}_{2L} = 0$ and $C^{\rm (S)}_{2L} = 1$, respectively. 
Because the atomic-limit ($t=0$) ground state is a simple, non-degenerate, and gapped product state, it must evolve continuously into the $J \to \infty$ ground state (except in the regime $m \pm J/2 = \ca O(t)$). 
Hence, phase~IV must carry $C^{\rm (S)}_{2L} = 1$, while the weak-disorder trivial phase (phase~I) has $C^{\rm (S)}_{2L} = 0$.
Thus, phases~I and~IV are topologically distinct and cannot be adiabatically connected.

Strategy (ii) proceeds with a completely different argument. 
Our TH calculations show that the transition between the $m<0$ ($C=+2$) and $m>0$ ($C=-2$) sectors of phase~III is not inherited from the clean model but is induced by {\em zeros} of the disorder-averaged Green’s function crossing the chemical potential. 
Indeed, the vertical $m=0$ line represents a conventional band-closure transition only for weak $J$.
At $J \approx 2.5$, it passes a tetra-critical point beyond which the transition is governed by Green's-function zeros.
Phase III (for $m<0$ or $m>0$) is topologically nontrivial but the topology is carried by Green's-function zeros.
A clean-limit Green's function, on the other hand, does not exhibit any zero.
Hence, phase III represents a novel phase that is {\em induced} by disorder and may thus be called a ``topological Anderson insulator.'' 

The mechanism leading to a topological Anderson insulator is reminiscent of the topological Mott insulator found in Ref.\ \cite{WCA+23}, where the topology is likewise encoded in zeros of the interacting Green’s function. 
Such states host boundary zeros that manifest only indirectly, for example, through annihilation with edge poles when placed in contact with a conventional topological insulator. 
An analogous scenario is expected in the present strong-disorder regime, and numerical studies in this direction are underway.

Finally, while the phase diagrams obtained from the TBC and TH approaches agree quantitatively in almost all respects, they differ along the $m=0$ line beyond the tetra-critical point at $J \approx 2.5$. 
Here, the TBC data do not support any transition between the $m<0$ ($C=+2$) and $m>0$ ($C=-2$) phases. 
{\em A posteriori}, this is readily understood: the TBC method computes the Chern number from the projector onto occupied {\em states} and is therefore insensitive to zeros of the disorder-averaged Green's function. 
We thus arrive at another notable conclusion: although the TBC and TH invariants coincide in the clean limit, they must be regarded as distinct topological invariants in the presence of finite disorder.
\\

\acknowledgments
We would like to thank Henryk Behrens and David Kr\"uger for discussions.
This work was supported by the Deutsche Forschungsgemeinschaft (DFG, German Research Foundation) through the research unit QUAST, FOR 5249 (project P8), project ID 449872909, and through the Cluster of Excellence ``Advanced Imaging of Matter'' - EXC 2056, project ID 390715994. 

\section*{Data availability}

The data that support the findings of this article are openly available \cite{data}.

\end{document}